\documentclass[conference]{IEEEtran}
\IEEEoverridecommandlockouts
\usepackage[sort&compress,comma,numbers]{natbib}
\usepackage{soul}
\usepackage{amsmath,amssymb,amsfonts}
\usepackage{algorithmic}
\usepackage{graphicx}
\usepackage{hyperref}
\usepackage{textcomp}
\usepackage{xcolor}
 \usepackage{booktabs} 
 \usepackage{multirow}
 \usepackage{flushend}
\usepackage[font=small]{caption}
\usepackage[inkscapelatex=false]{svg}
\def\BibTeX{{\rm B\kern-.05em{\sc i\kern-.025em b}\kern-.08em
    T\kern-.1667em\lower.7ex\hbox{E}\kern-.125emX}}
\begin{document}

\renewcommand{\figureautorefname}{Fig.}
\renewcommand{\tableautorefname}{TABLE.}
\renewcommand{\sectionautorefname}{Sec.}
\renewcommand{\equationautorefname}{Eq.}

\title{Compressed Video Super-Resolution based on Hierarchical Encoding}

\author{
Yuxuan Jiang$^1$,
Siyue Teng$^1$
Qiang Zhu$^{2,1}$,
Chen Feng$^1$,
Chengxi Zeng$^1$,\\
Fan Zhang$^1$,
Shuyuan Zhu$^2$,
Bing Zeng$^2$,
and David Bull$^1$ \\
\textit{$^1$ University of Bristol, $^2$ University of Electronic Science and Technology of China} \\
$^1$ \textit{\{yuxuan.jiang, siyue.teng, chen.feng, simon.zeng, fan.zhang, dave.bull\}@bristol.ac.uk}\\
$^2$ \textit{\{zhuqiang@std., eezsy@, eezeng@\}uestc.edu.cn}\\
}

\maketitle

\begin{abstract}
This paper presents a general-purpose video super-resolution (VSR) method, dubbed VSR-HE, specifically designed to enhance the perceptual quality of compressed content. Targeting scenarios characterized by heavy compression, the method upscales low-resolution videos by a ratio of four, from 180p to 720p or from 270p to 1080p. VSR-HE adopts hierarchical encoding transformer blocks and has been sophisticatedly optimized to eliminate a wide range of compression artifacts commonly introduced by H.265/HEVC encoding across various quantization parameter (QP) levels. To ensure robustness and generalization, the model is trained and evaluated under diverse compression settings, allowing it to effectively restore fine-grained details and preserve visual fidelity. The proposed VSR-HE has been officially submitted to the ICME 2025 Grand Challenge on VSR for Video Conferencing (Team BVI-VSR), under both the Track 1 (General-Purpose Real-World Video Content) and Track 2 (Talking Head Videos).
\end{abstract}


\section{Introduction}

Visual content plays an increasingly important role in our current digital ecosystem. With the proliferation of smartphones, tablets, and other digital devices, the consumption of video content has surged across a wide range of applications, including live streaming, digital broadcasting, video conferencing, and intelligent surveillance. These video-centric services now account for a dominant share — approximately 80\% — of global internet traffic, as reported by Cisco \cite{r:cisco2}.

At the core of enabling efficient video transmission lies video compression, a fundamental and long-standing research area in image and video processing. Its role is vital in balancing the trade-off between the high bitrate required to preserve rich, immersive video content and the limited bandwidth resources typically available in real-world scenarios. Over the past decades, the field has seen remarkable progress, leading to the development of cutting-edge video coding standards such as H.265/HEVC \cite{h265HEVC}, H.266/VVC \cite{VVCSoftware_VTM} and AOMedia (AOM) AV1 \cite{AV1}. Despite these advancements, current video codecs still rely on traditional rate-distortion optimization frameworks inherited from predecessors like HEVC and VP9 \cite{VP9}. More recently, both MPEG and AOM have initiated the exploration of new video coding algorithms beyond their latest standards, with working codecs ECM (Enhanced Compression Model) \cite{ecm_github} and AVM (AOM Video Model) \cite{avm_github}, respectively. However, current solutions in these models are based on their legacy foundations, which may limit their ability to fully meet the rapidly escalating demands of next-generation media applications, especially when operating at ultra-high spatial resolutions and maintaining a delicate balance between encoding/decoding efficiency and overall performance \cite{teng2024benchmarking}.

\begin{figure}[t]
    \centering
    \includegraphics[width=1\linewidth]{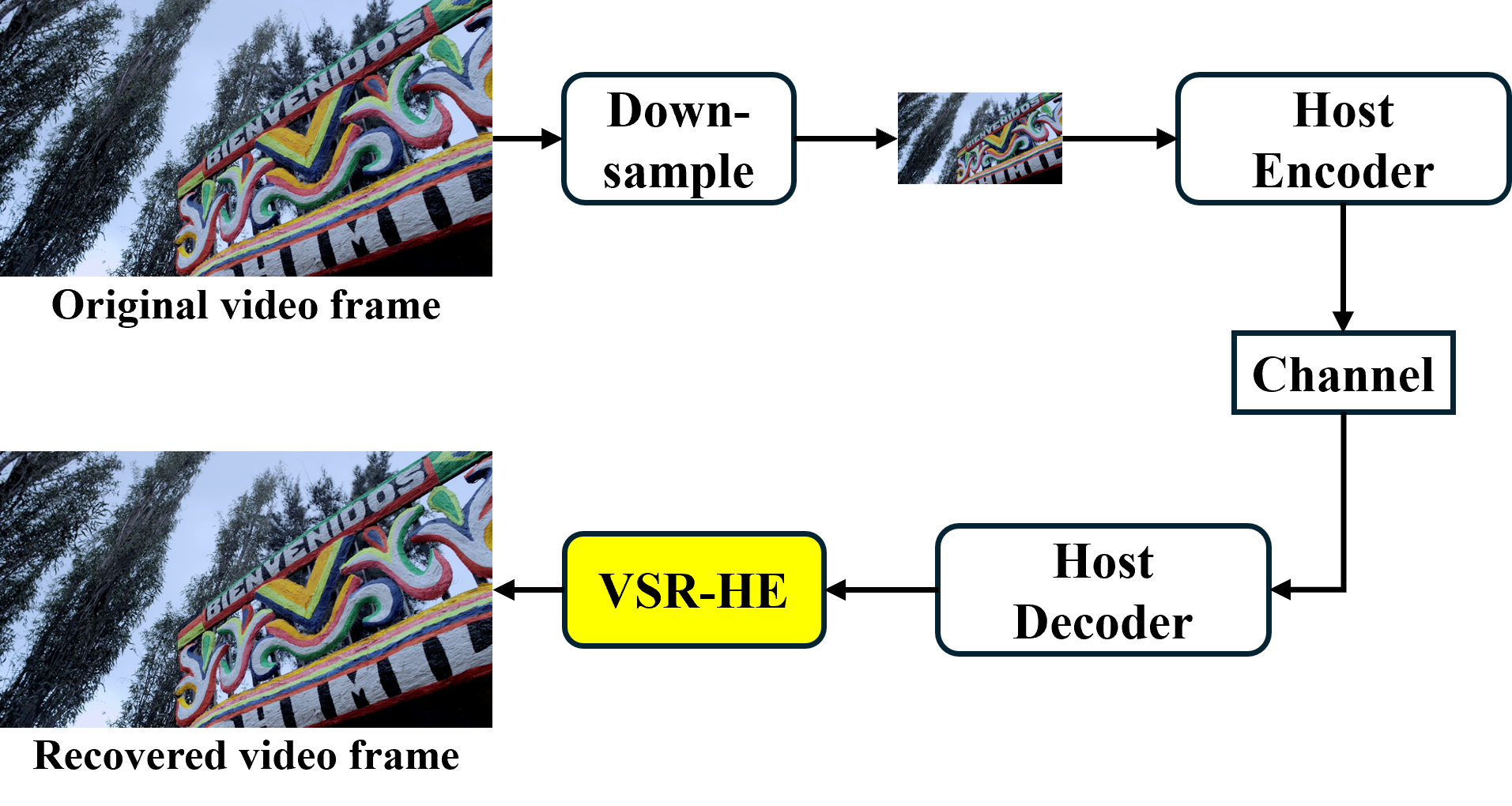}
    \caption{The applied coding framework, with a VSR-HE module serving as SR.}
    \label{fig:coding_frame}
\end{figure}

To address these limitations, deep learning has emerged as a transformative tool for video compression. Inspired by the advances of image super-resolution \cite{jiang2024mtkd, liang2021swinir, jiang2024hiif, jiang2025c2d, zhu2025blind, lim2017enhanced, peng2025instance}, a growing body of learning-based VSR coding approaches has been proposed in recent years \cite{yan2018convolutional, zhang2020enhancing, ma2020cvegan, ma2020mfrnet, jiang2024rtsr, jiang2024compressing, feng2022vistra3, conde2024aim, zhu2024cpga, zhu2025fcvsr}, demonstrating impressive improvements in coding efficiency when integrated into standard video codecs. It is noted that, however, most CNN-based video compression techniques are trained using simple distortion-based loss functions, such as mean squared error (MSE) or L1 loss. While these metrics are easy to compute, they correlate poorly with human visual perception and often result in suboptimal perceptual quality \cite{ma2020cvegan}.

\begin{figure*}[t]
    \centering
    \includegraphics[width=1\linewidth, trim=60 0 0 50]{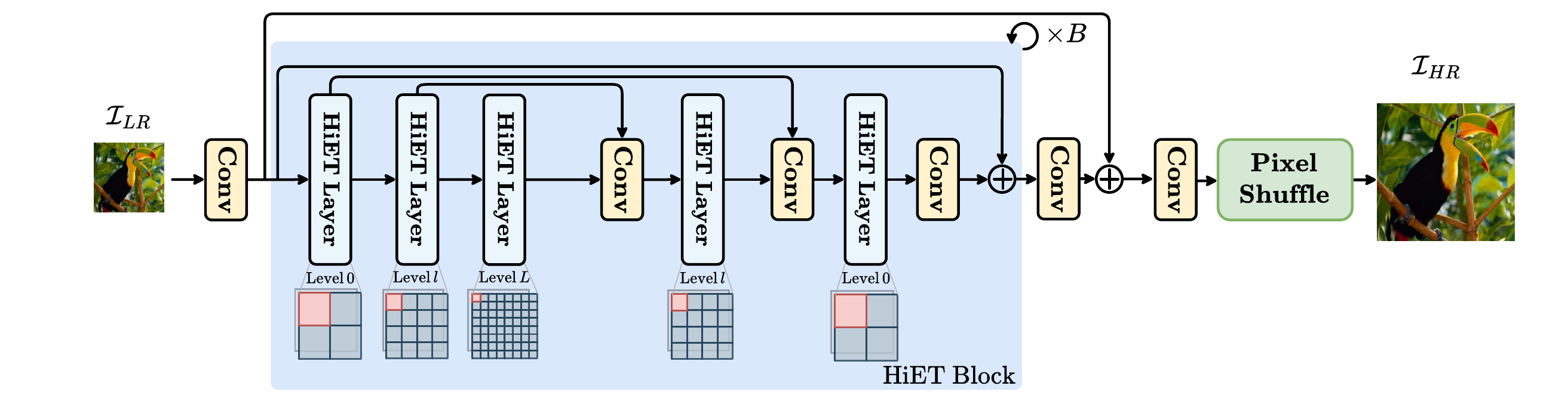}
    \caption{The architecture of the proposed network architecture for super resolution. The HiET layers are adopted from \cite{jiang2025c2d}. Window sizes are set to [64, 32, 8, 32, 64], with $B=6$ and the channel dimension of 126.}
    \label{fig:netwoork}
\end{figure*}

\begin{figure*}[t]
        \scriptsize
    \centering
    \begin{minipage}[b]{0.325\linewidth}
        \centering
        \centerline{\includegraphics[width=.98\linewidth]{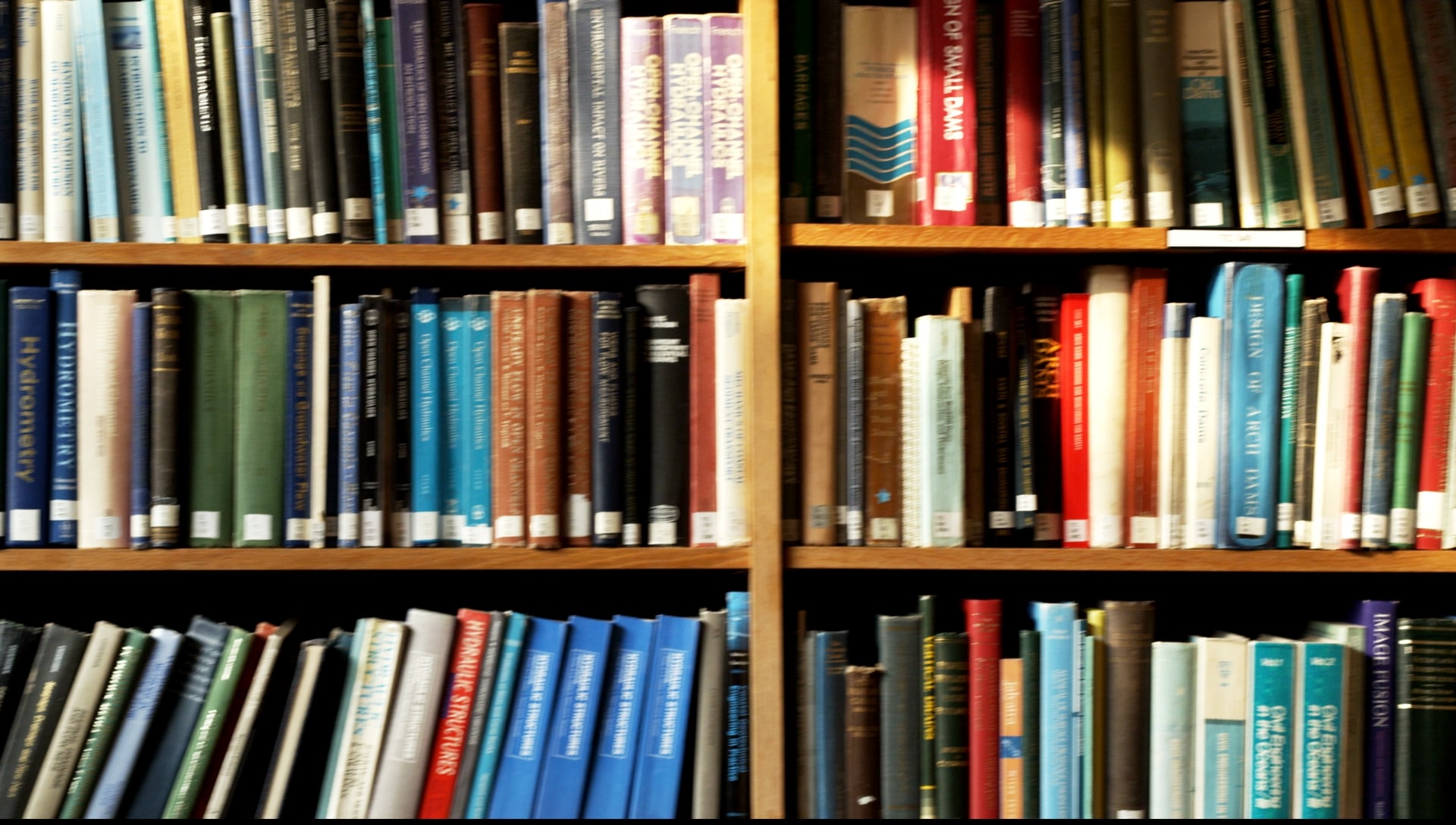}}
        \text{BBookcaseBVITexture}\vspace{.1cm}
    \end{minipage}
    \begin{minipage}[b]{0.325\linewidth}
        \centering
        \centerline{\includegraphics[width=.98\linewidth]{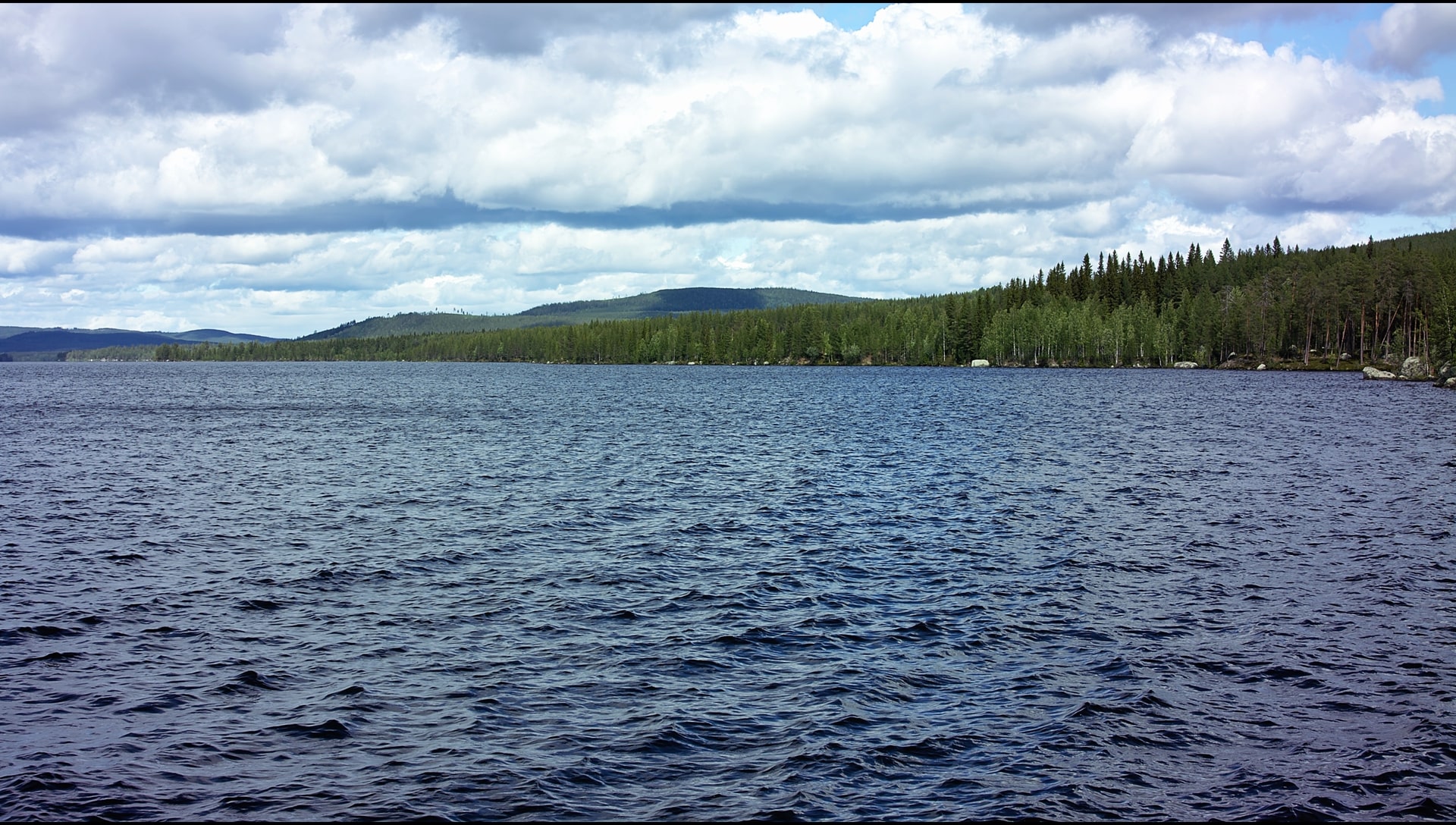}}
        \text{BForestLake}\vspace{.1cm}
            \end{minipage}
    \begin{minipage}[b]{0.325\linewidth}
        \centering
        \centerline{\includegraphics[width=.98\linewidth]{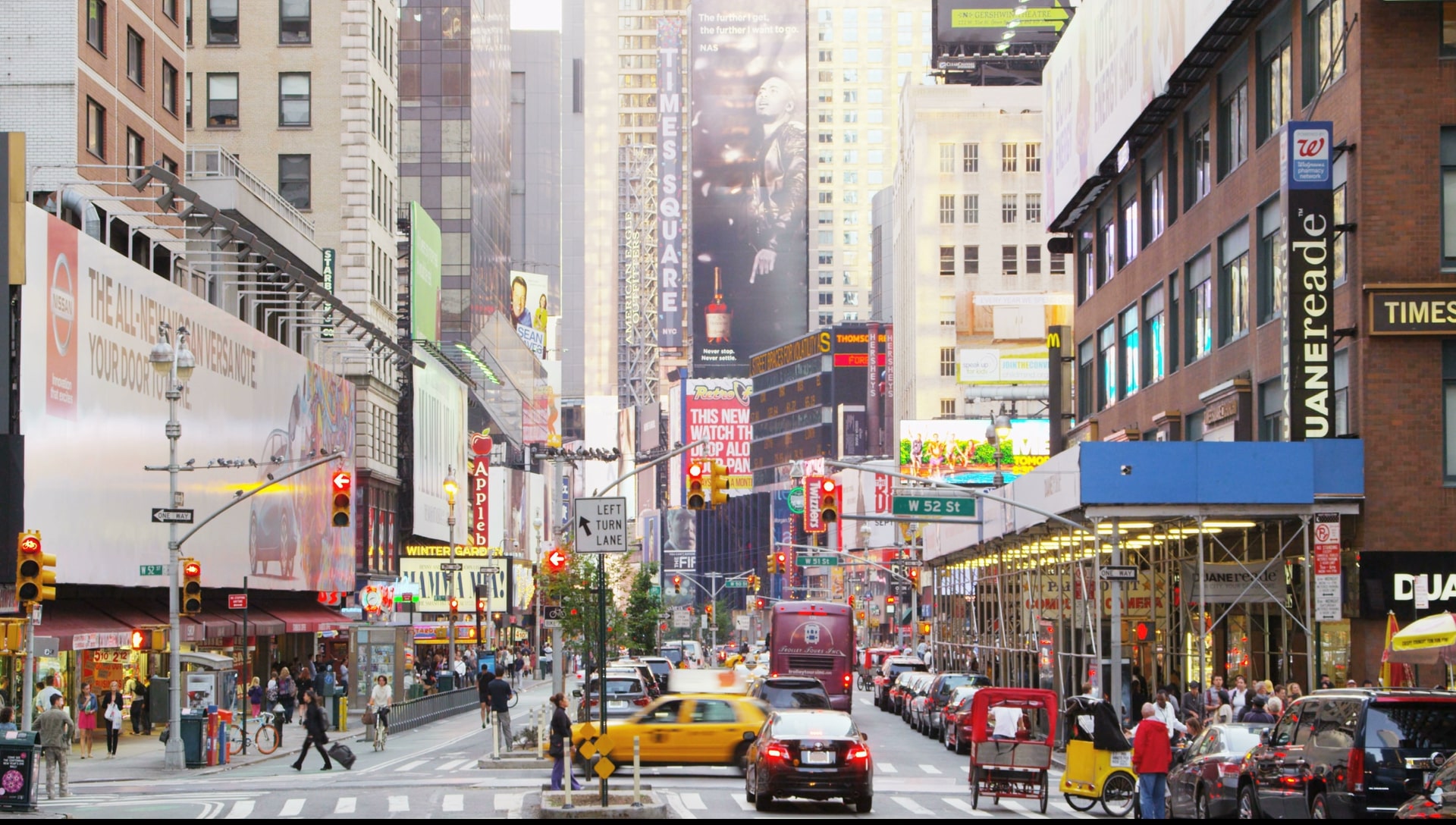}}
        \text{BNewYorkStreetDareful}\vspace{.1cm}
        \end{minipage}

    \centering
    \begin{minipage}[b]{0.325\linewidth}
        \centering
        \centerline{\includegraphics[width=.98\linewidth]{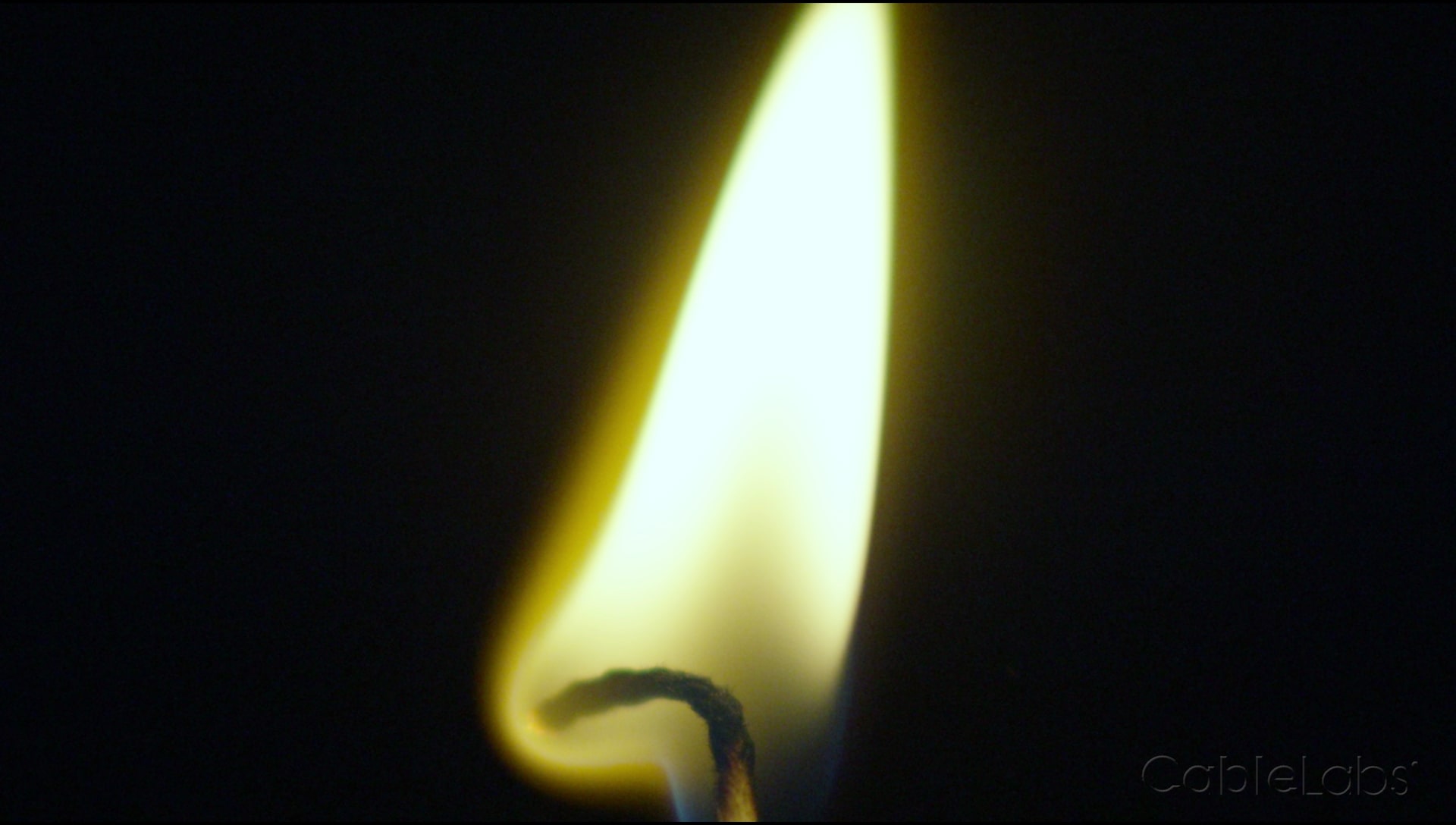}}
        \text{BAncientThoughtS14}\vspace{.1cm}
            \end{minipage}
    \begin{minipage}[b]{0.325\linewidth}
        \centering
        \centerline{\includegraphics[width=.98\linewidth]{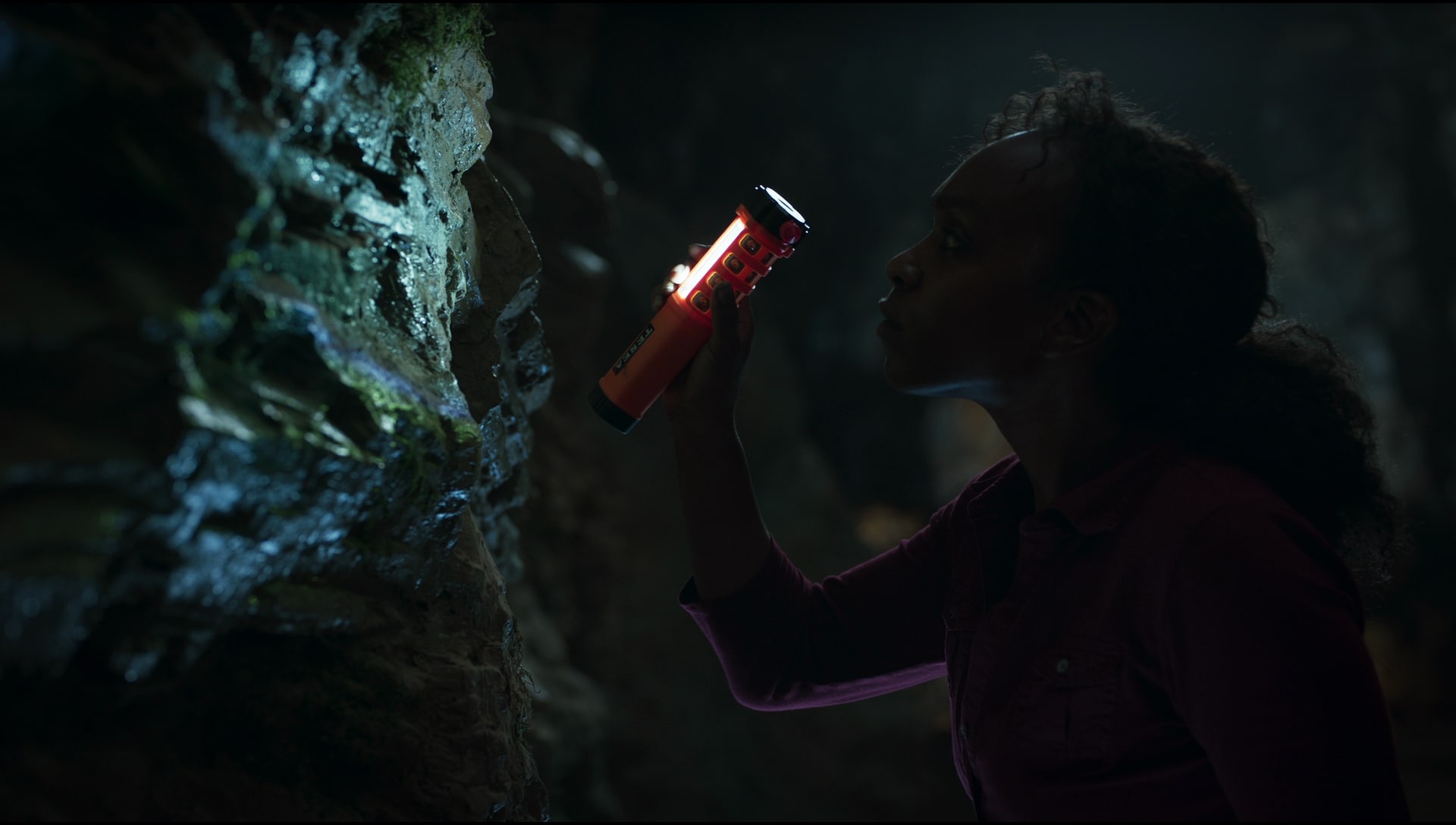}}
        \text{BAscStem2S4}\vspace{.1cm}
    \end{minipage}
        \begin{minipage}[b]{0.325\linewidth}
        \centering
        \centerline{\includegraphics[width=.98\linewidth]{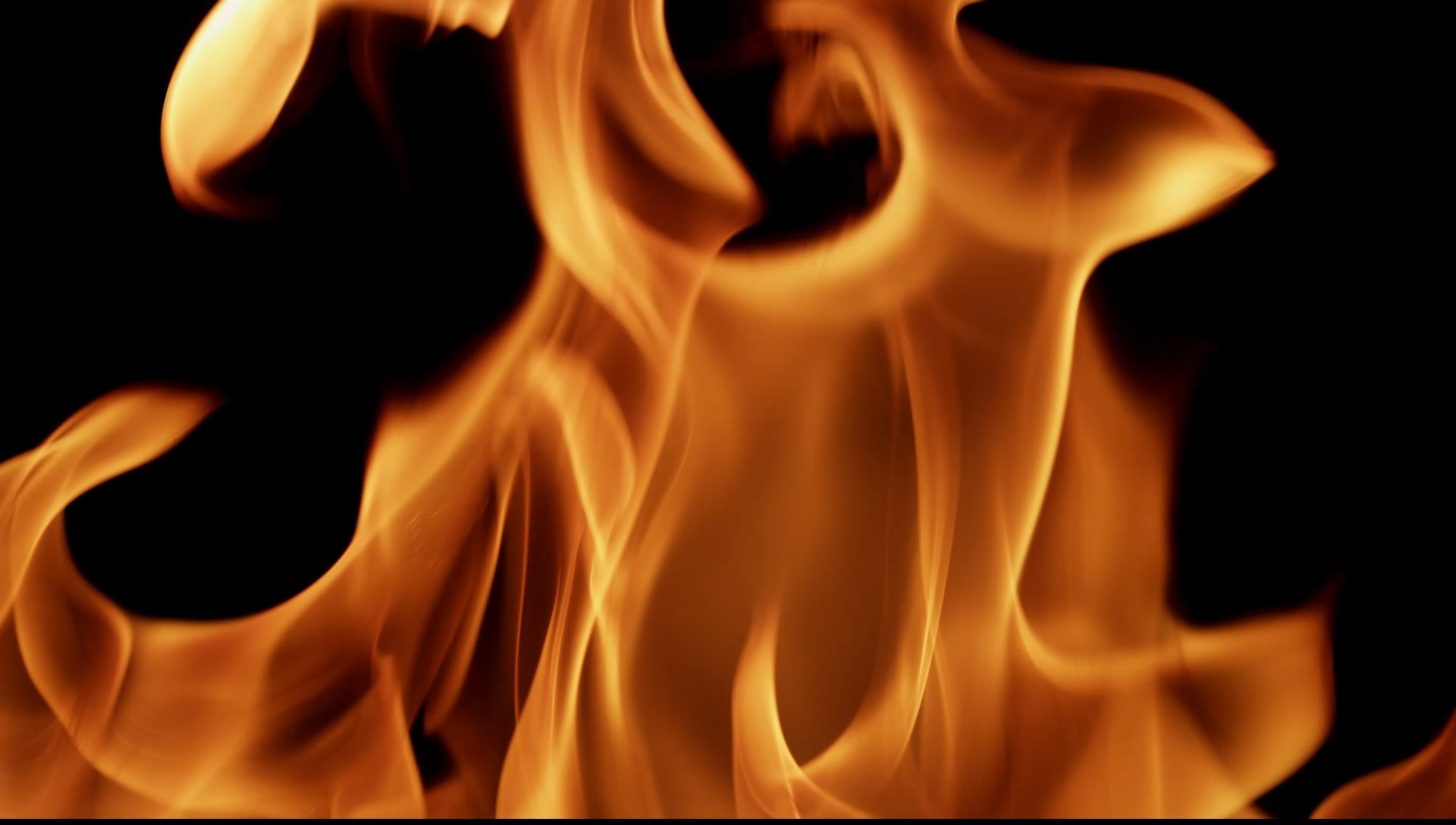}}
        \text{BFireS18Mitch}\vspace{.1cm}
    \end{minipage}
    
    \caption{Sequence thumbnails of training content from BVI-AOM \cite{nawala2024bvi} dataset.}
    \label{fig:bviexample}
\end{figure*}

In this paper, we propose a deep learning-based video super-resolution approach, which has been submitted to the ICME 2025 Grand Challenge on Video Super-Resolution for Video Conferencing (Track 1 \& 2). The proposed method builds upon a previously developed efficient architecture, the HiET block \cite{jiang2025c2d}, and employs a perceptual loss function (PLF) combined with GAN-based training, inspired by the CVEGAN framework \cite{ma2020cvegan}. In addition to the training set provided by the organizers, we also used the BVI-AOM \cite{nawala2024bvi} database as a supplement to further improve the model generalization and boost the performance. In accordance with the challenge requirements, our method strictly processes each frame independently during upscaling and enhancement. This approach, denoted as VSR-HE, has been evaluated on H.265/HEVC compressed content (ICME Grand Challenge validation video sequences) and demonstrates consistent improvements across multiple evaluation metrics—including PSNR, SSIM, MS-SSIM, and VMAF. Notably, it outperforms both conventional upscaling methods such as bicubic filter and recent learning-based VSR models such as EDSR \cite{lim2017enhanced} and SwinIR \cite{liang2021swinir}.

The rest of the paper is organized as follows. Section \ref{sec:PA} describes the proposed VSR-HE method, the integrated coding framework, and the training process. The coding results are then presented in Section \ref{sec:RD}. Finally, Section \ref{sec:c} concludes the paper and outlines the future work.

\section{Proposed Algorithm}
\label{sec:PA}

The coding framework is illustrated in \autoref{fig:coding_frame}. Prior to encoding, the original YUV 720p videos are downsampled by a factor of 4 using a bicubic filter for Track 1, whereas the original YUV 1080p videos are downsampled by a factor of 4 using the Lanczos filter for Track 2. A H.265/HEVC video codec \cite{h265HEVC} serves as the Host Encoder to compress the low-resolution video in a low-power setting tailored to low-delay conferencing scenarios. At the decoder, when the low-resolution video stream is decoded, the proposed VSR-HE approach is applied to reconstruct the full-resolution video content. Details regarding the network architecture and the training process are described below.


\begin{figure*}[t]
        \scriptsize
    \centering
    \begin{minipage}[b]{0.325\linewidth}
        \centering
        \centerline{\includegraphics[width=.98\linewidth]{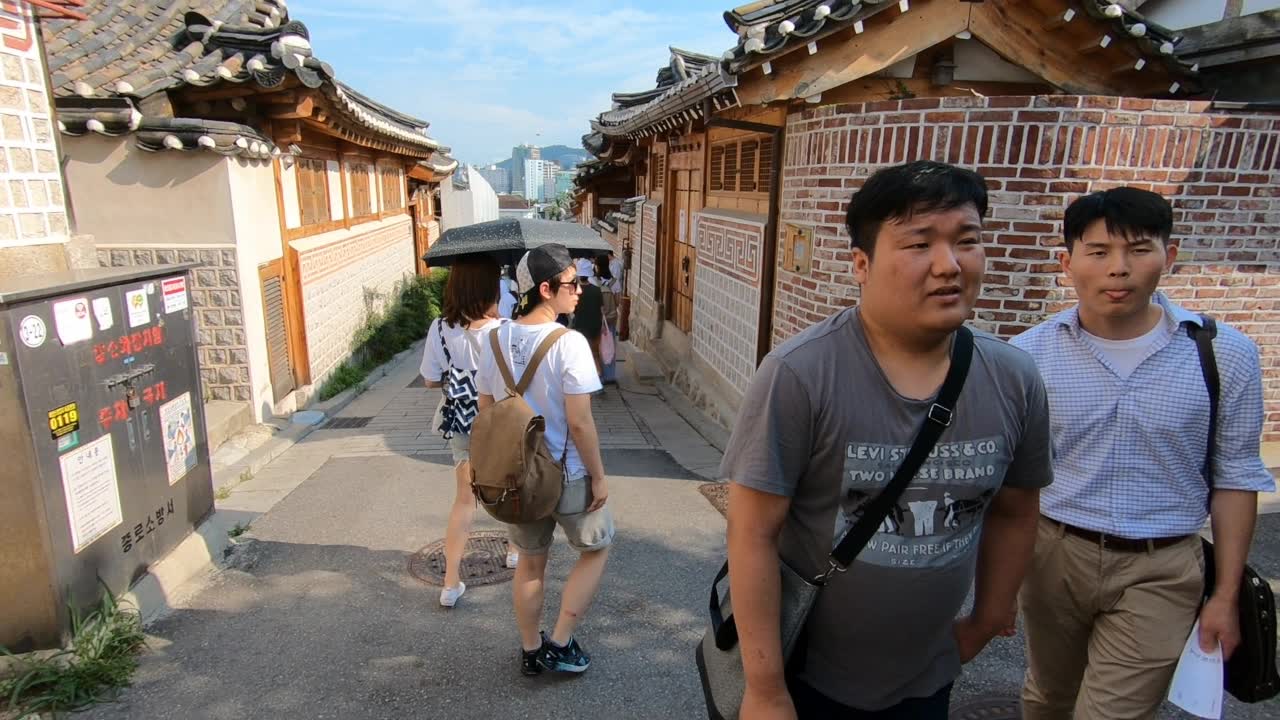}}
        \text{Sequence1 GT}\vspace{.1cm}
        \end{minipage}
    \begin{minipage}[b]{0.325\linewidth}
        \centering
        \centerline{\includegraphics[width=.98\linewidth]{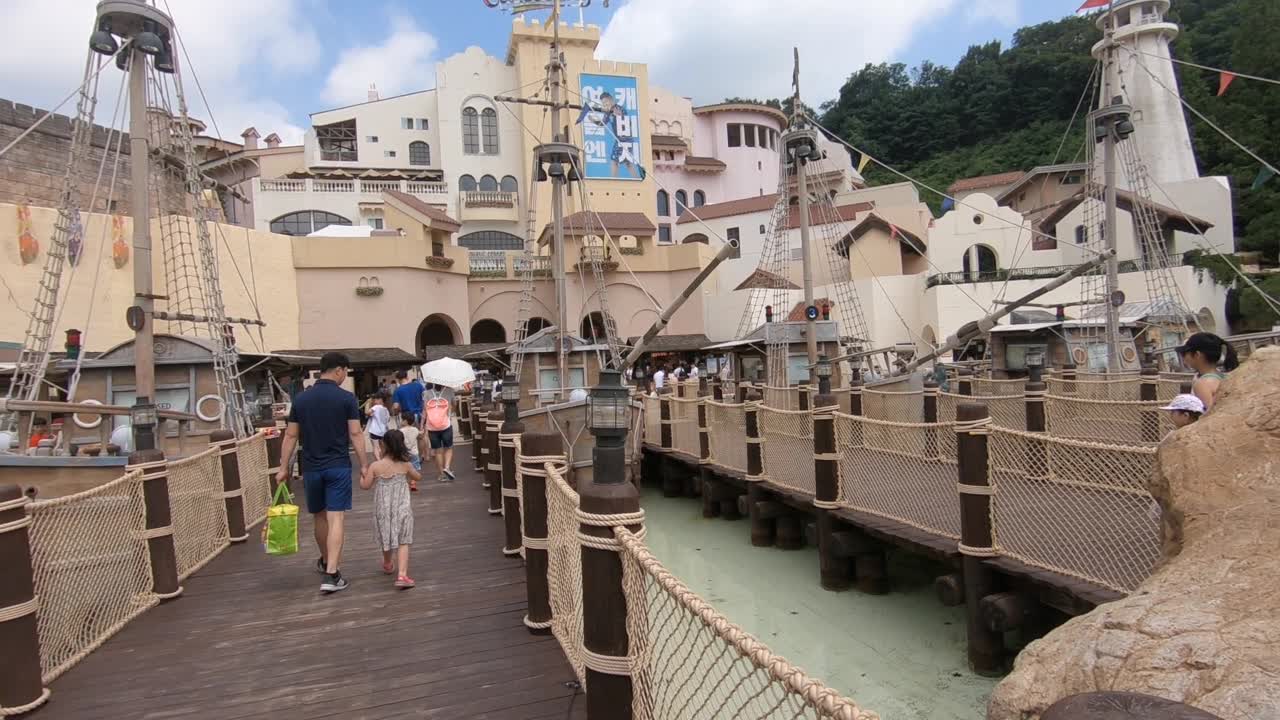}}
        \text{Sequence2 GT}\vspace{.1cm}
            \end{minipage}
    \begin{minipage}[b]{0.325\linewidth}
        \centering
        \centerline{\includegraphics[width=.98\linewidth]{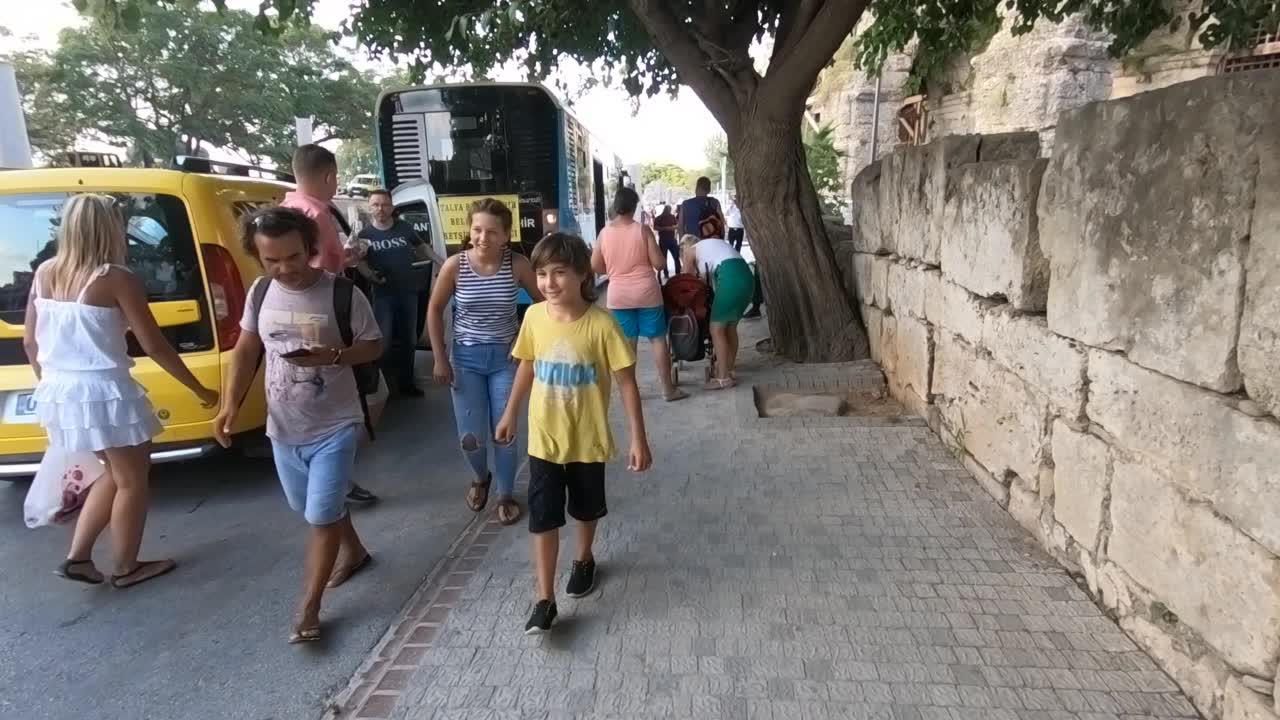}}
        \text{Sequence3 GT}\vspace{.1cm}
            \end{minipage}

    \begin{minipage}[b]{0.325\linewidth}
        \centering
        \centerline{\includegraphics[width=.98\linewidth]{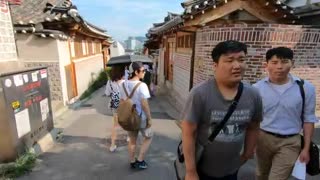}}
        \text{Sequence1 LR}\vspace{.1cm}
        \end{minipage}
    \begin{minipage}[b]{0.325\linewidth}
        \centering
        \centerline{\includegraphics[width=.98\linewidth]{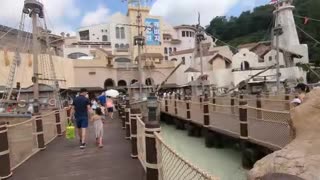}}
        \text{Sequence2 LR}\vspace{.1cm}
            \end{minipage}
    \begin{minipage}[b]{0.325\linewidth}
        \centering
        \centerline{\includegraphics[width=.98\linewidth]{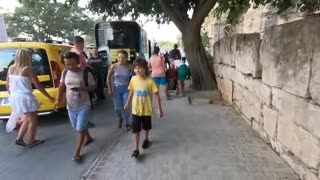}}
        \text{Sequence3 LR}\vspace{.1cm}
            \end{minipage}

    \begin{minipage}[b]{0.325\linewidth}
        \centering
        \centerline{\includegraphics[width=.98\linewidth]{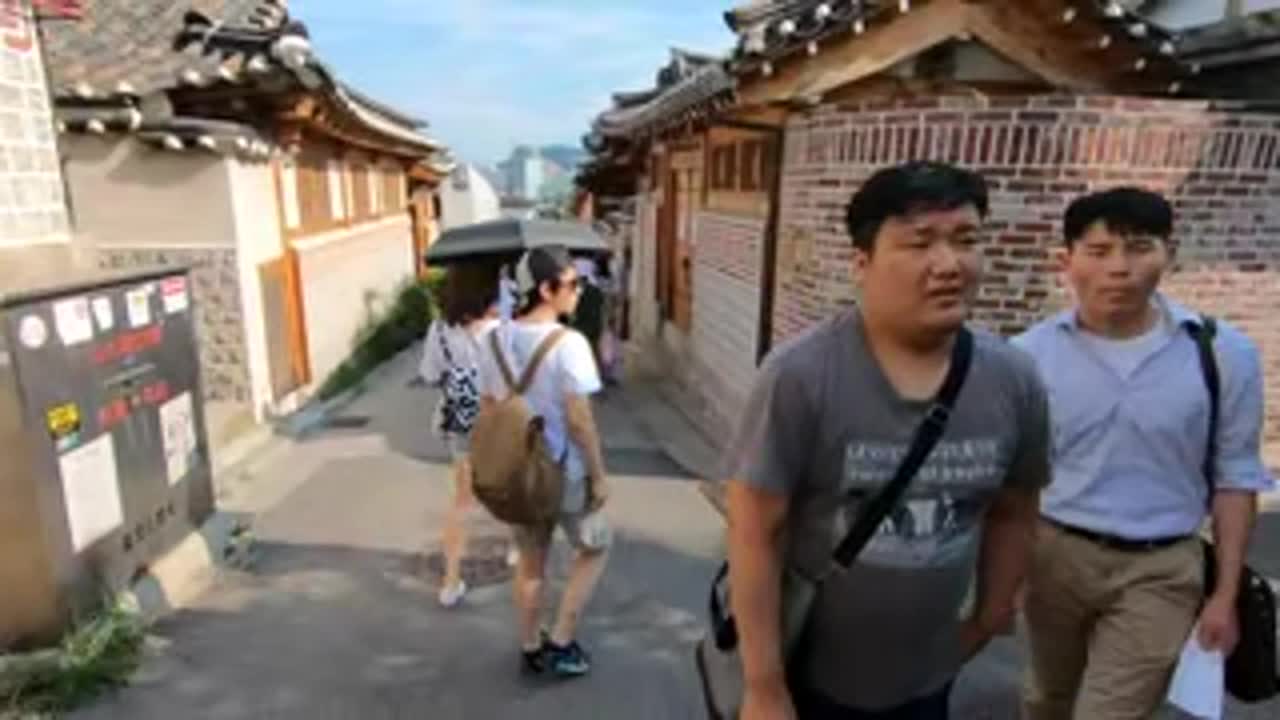}}
        \text{Sequence1 Bicubic}\vspace{.1cm}
        \end{minipage}
    \begin{minipage}[b]{0.325\linewidth}
        \centering
        \centerline{\includegraphics[width=.98\linewidth]{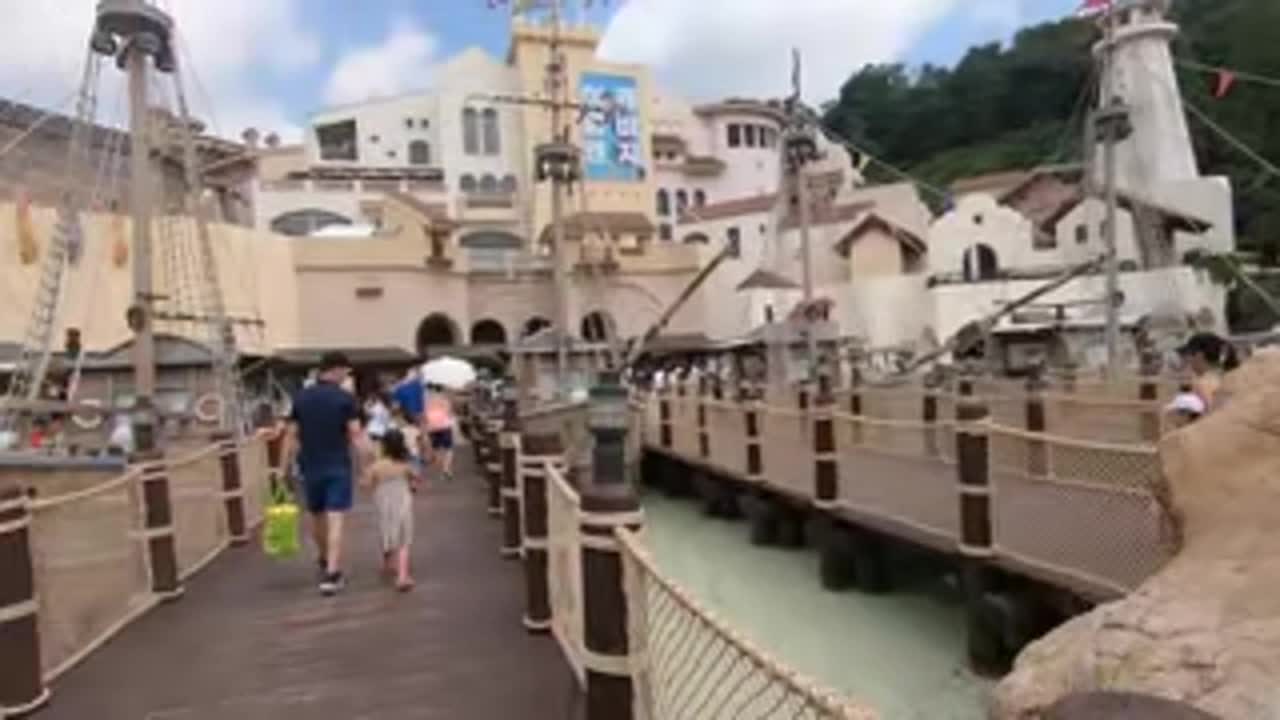}}
        \text{Sequence2 Bicubic}\vspace{.1cm}
            \end{minipage}
    \begin{minipage}[b]{0.325\linewidth}
        \centering
        \centerline{\includegraphics[width=.98\linewidth]{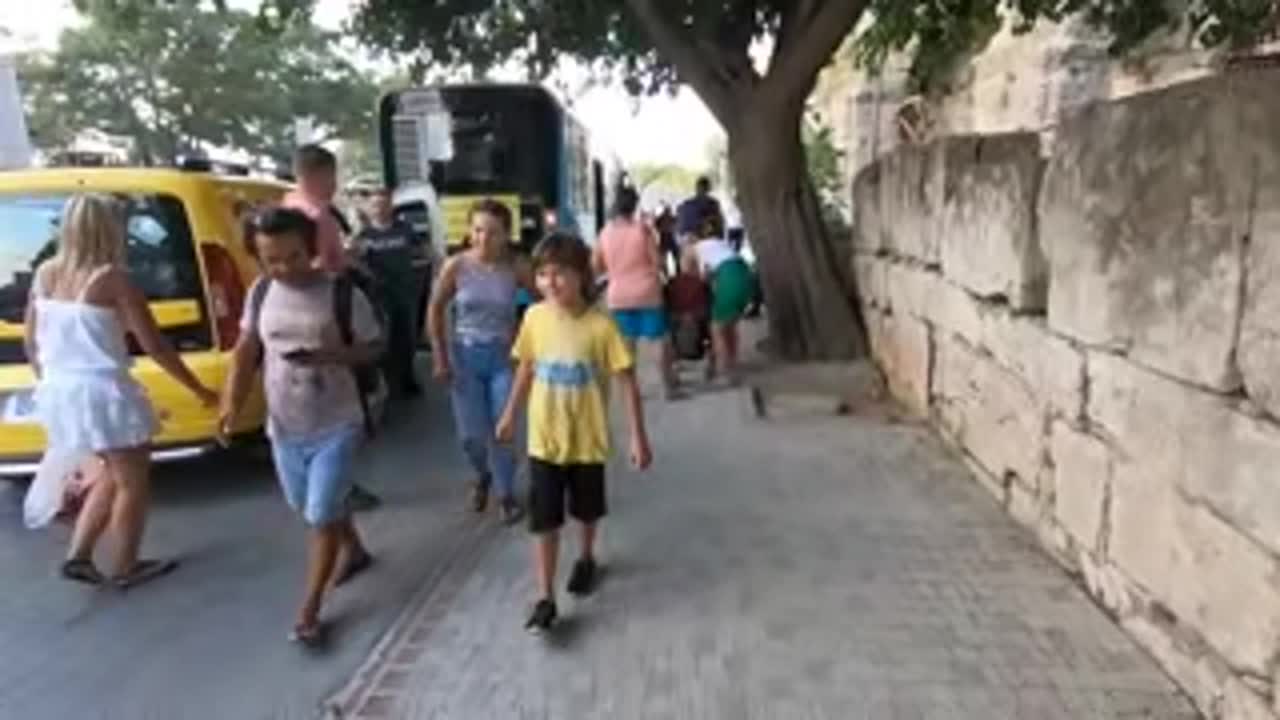}}
        \text{Sequence3 Bicubic}\vspace{.1cm}
            \end{minipage}

    \begin{minipage}[b]{0.325\linewidth}
        \centering
        \centerline{\includegraphics[width=.98\linewidth]{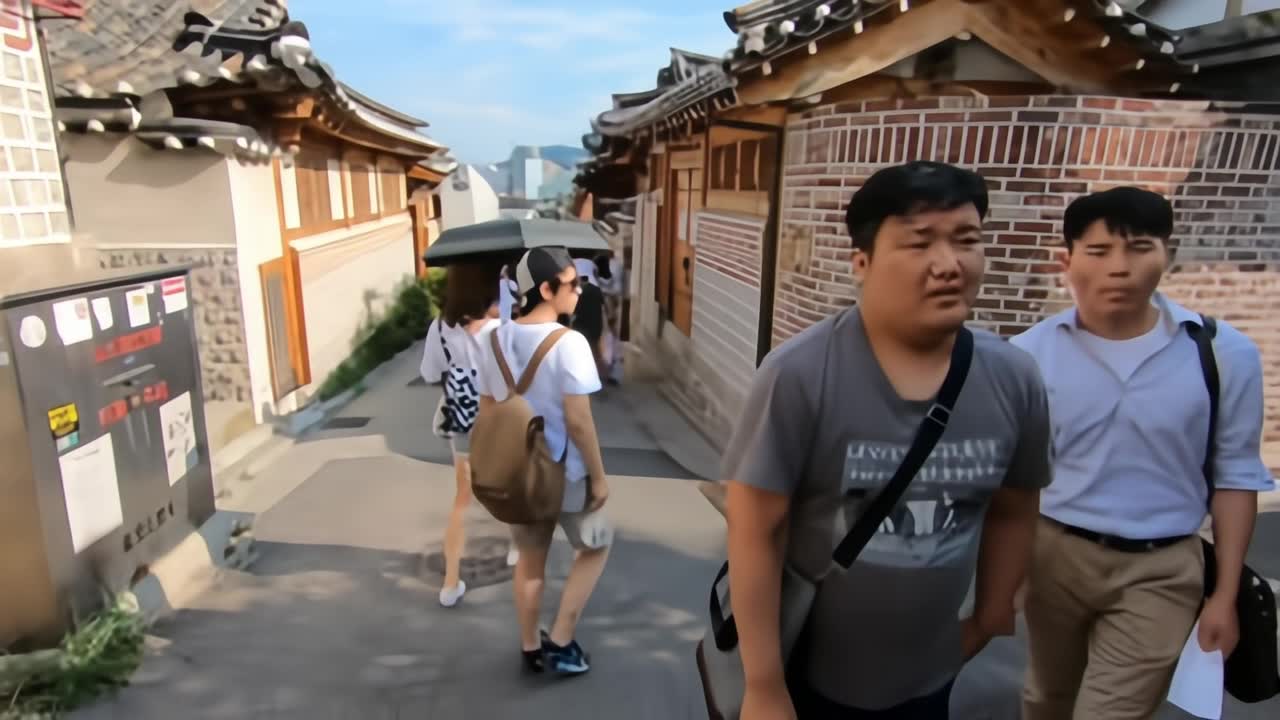}}
        \text{Sequence1 Ours}\vspace{.1cm}
        \end{minipage}
    \begin{minipage}[b]{0.325\linewidth}
        \centering
        \centerline{\includegraphics[width=.98\linewidth]{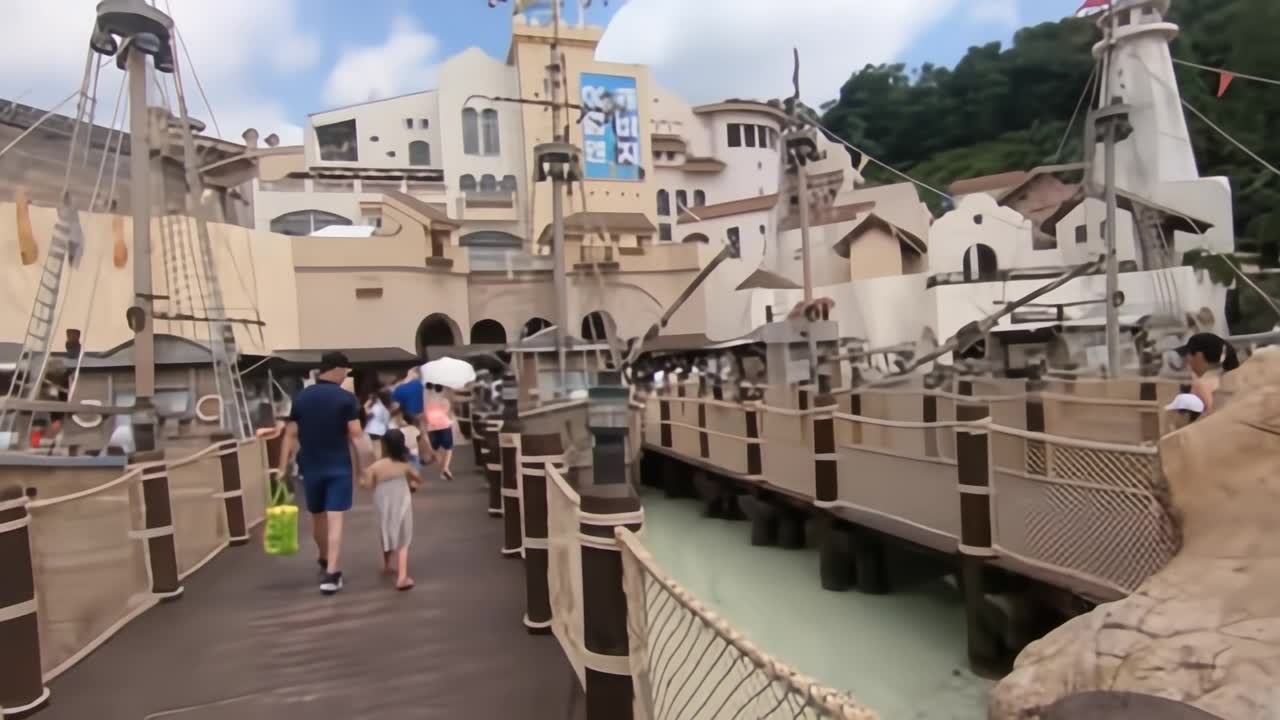}}
        \text{Sequence2 Ours}\vspace{.1cm}
            \end{minipage}
    \begin{minipage}[b]{0.325\linewidth}
        \centering
        \centerline{\includegraphics[width=.98\linewidth]{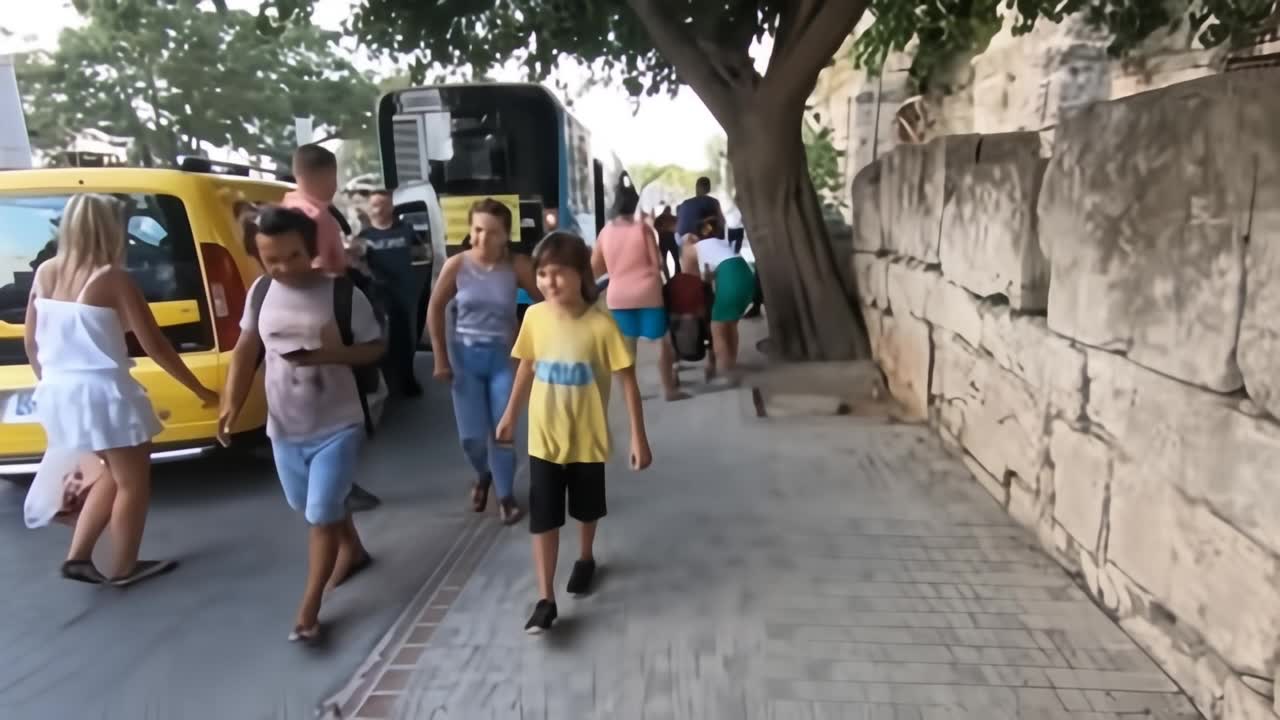}}
        \text{Sequence3 Ours}\vspace{.1cm}
            \end{minipage}
    
    
    \caption{Visual comparison of track1 SR reconstruction results.}
    \label{fig:example1}
\end{figure*}

\begin{figure*}[t]
        \scriptsize
    \centering
    \begin{minipage}[b]{0.325\linewidth}
        \centering
        \centerline{\includegraphics[width=.98\linewidth]{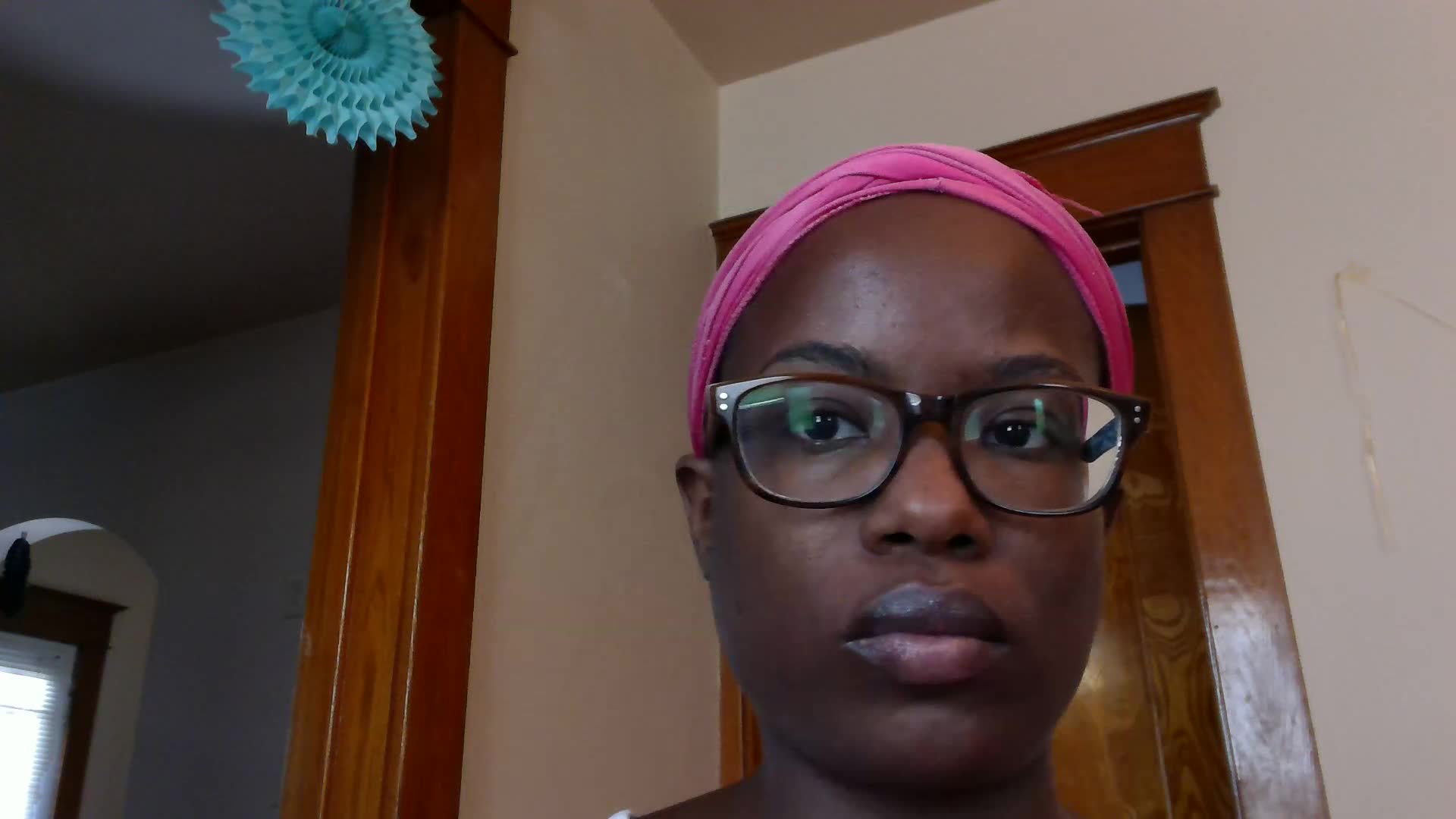}}
        \text{Sequence1 GT}\vspace{.1cm}
        \end{minipage}
    \begin{minipage}[b]{0.325\linewidth}
        \centering
        \centerline{\includegraphics[width=.98\linewidth]{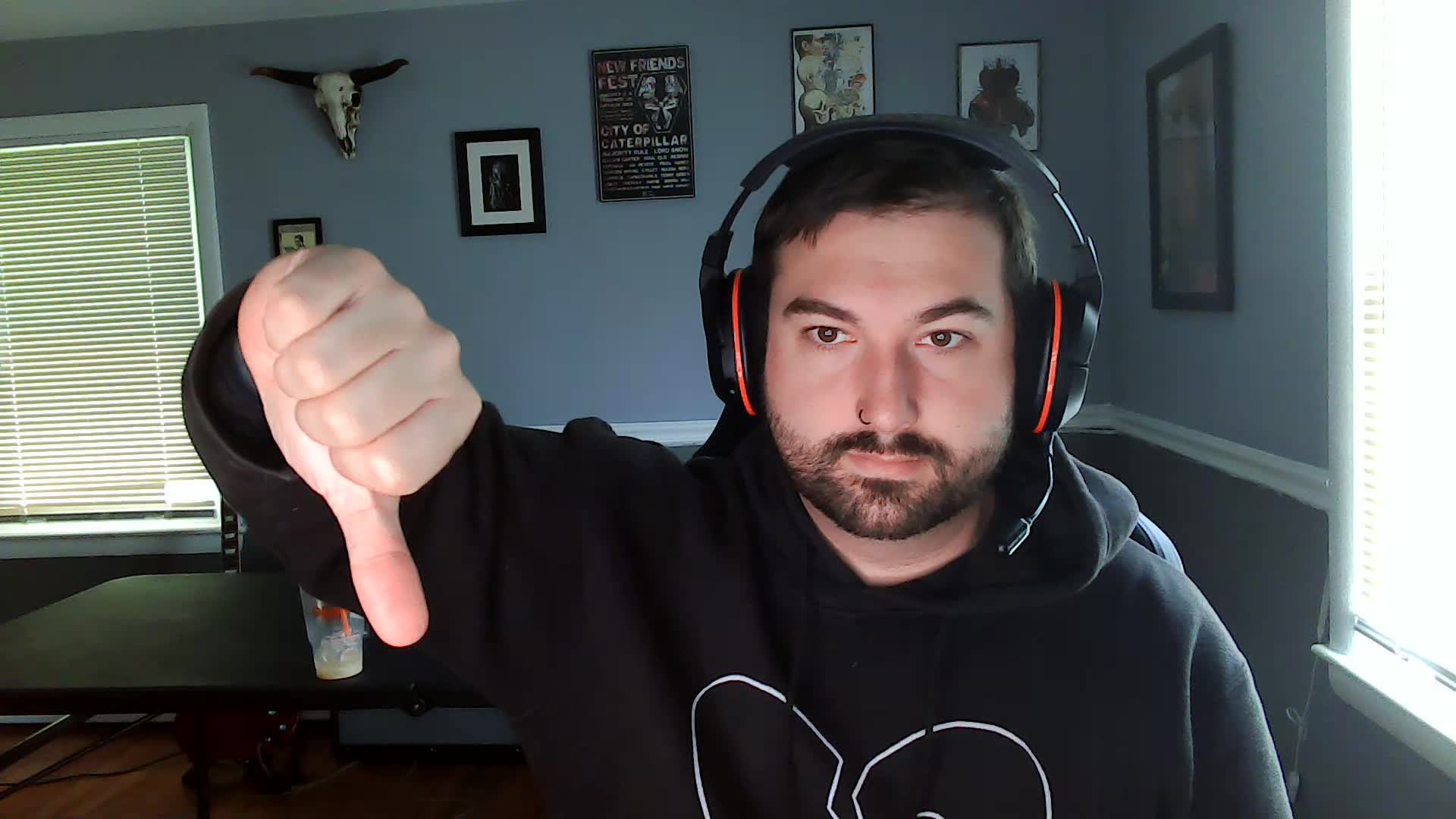}}
        \text{Sequence2 GT}\vspace{.1cm}
            \end{minipage}
    \begin{minipage}[b]{0.325\linewidth}
        \centering
        \centerline{\includegraphics[width=.98\linewidth]{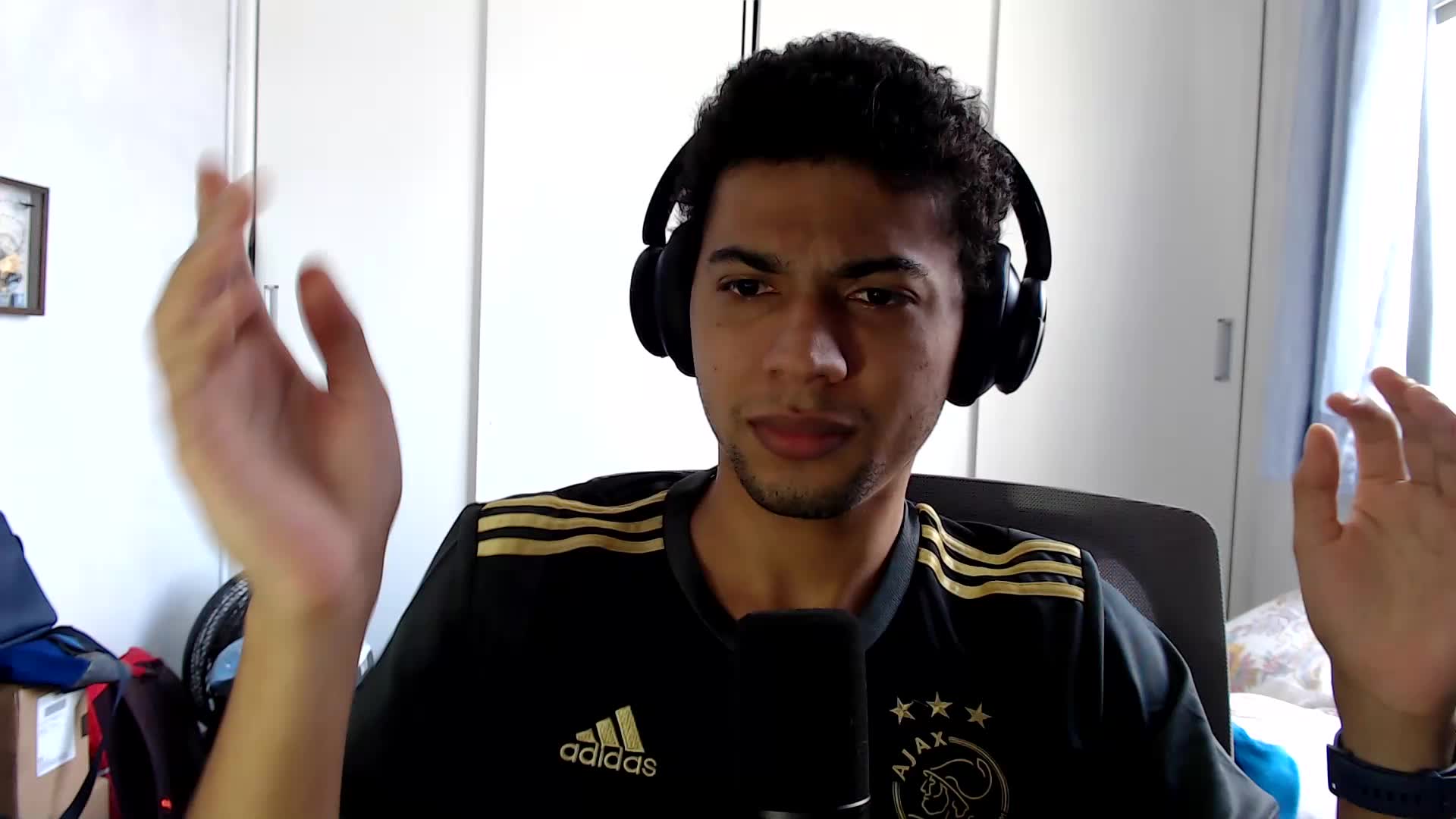}}
        \text{Sequence3 GT}\vspace{.1cm}
            \end{minipage}

    \begin{minipage}[b]{0.325\linewidth}
        \centering
        \centerline{\includegraphics[width=.98\linewidth]{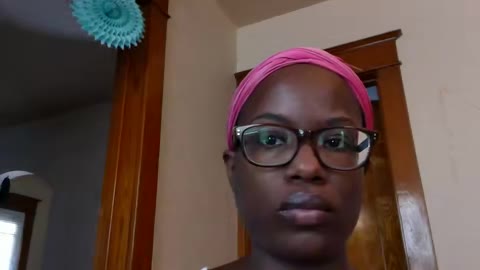}}
        \text{Sequence1 LR}\vspace{.1cm}
        \end{minipage}
    \begin{minipage}[b]{0.325\linewidth}
        \centering
        \centerline{\includegraphics[width=.98\linewidth]{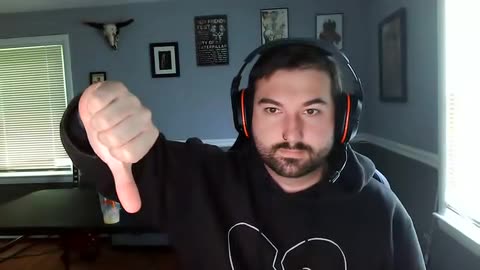}}
        \text{Sequence2 LR}\vspace{.1cm}
            \end{minipage}
    \begin{minipage}[b]{0.325\linewidth}
        \centering
        \centerline{\includegraphics[width=.98\linewidth]{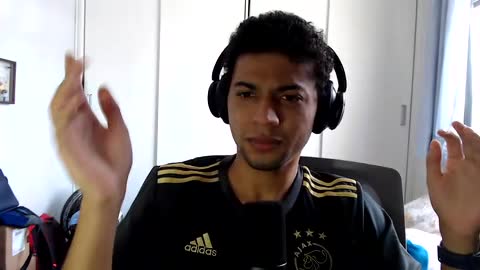}}
        \text{Sequence3 LR}\vspace{.1cm}
            \end{minipage}

    \begin{minipage}[b]{0.325\linewidth}
        \centering
        \centerline{\includegraphics[width=.98\linewidth]{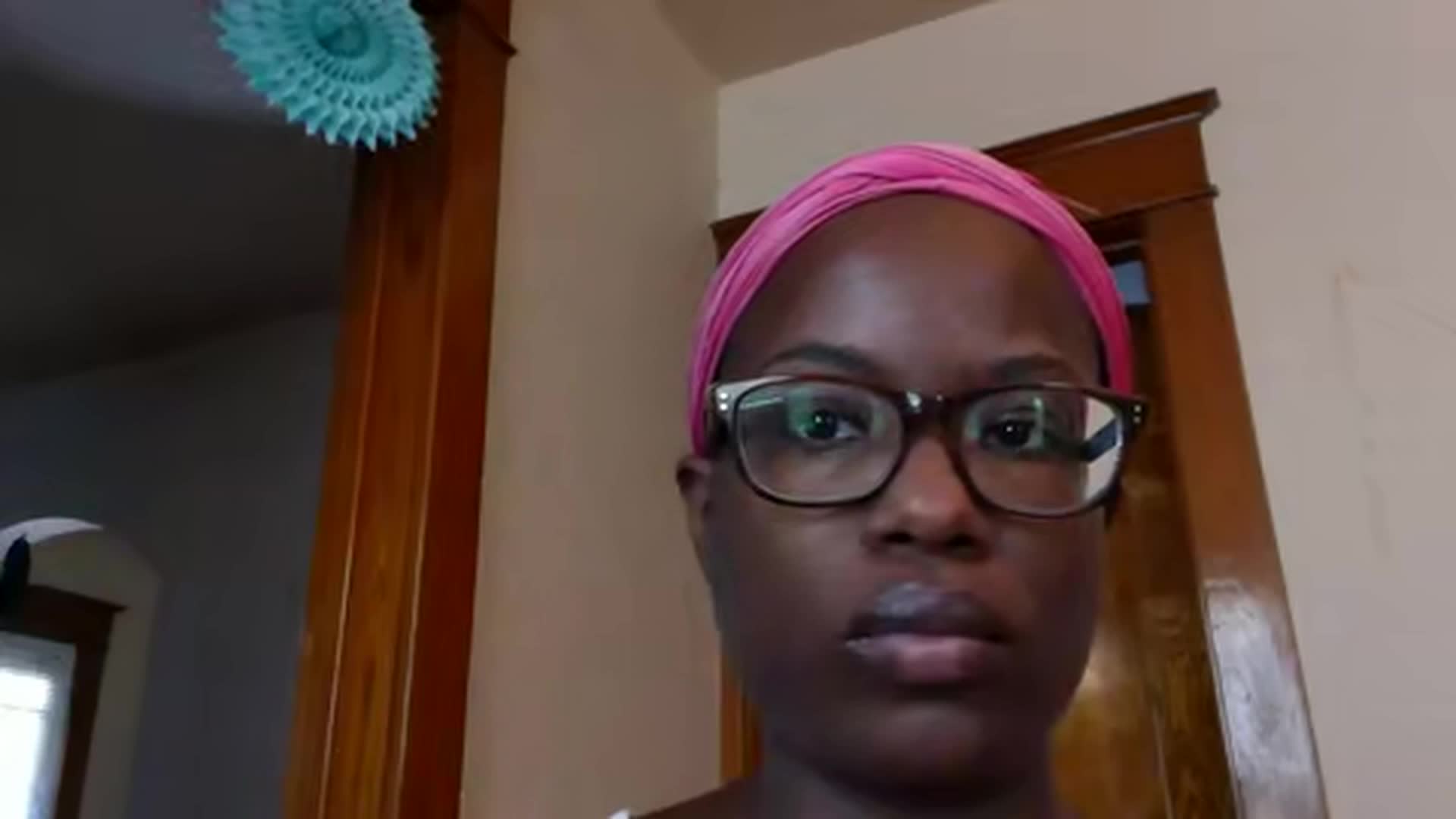}}
        \text{Sequence1 Bicubic}\vspace{.1cm}
        \end{minipage}
    \begin{minipage}[b]{0.325\linewidth}
        \centering
        \centerline{\includegraphics[width=.98\linewidth]{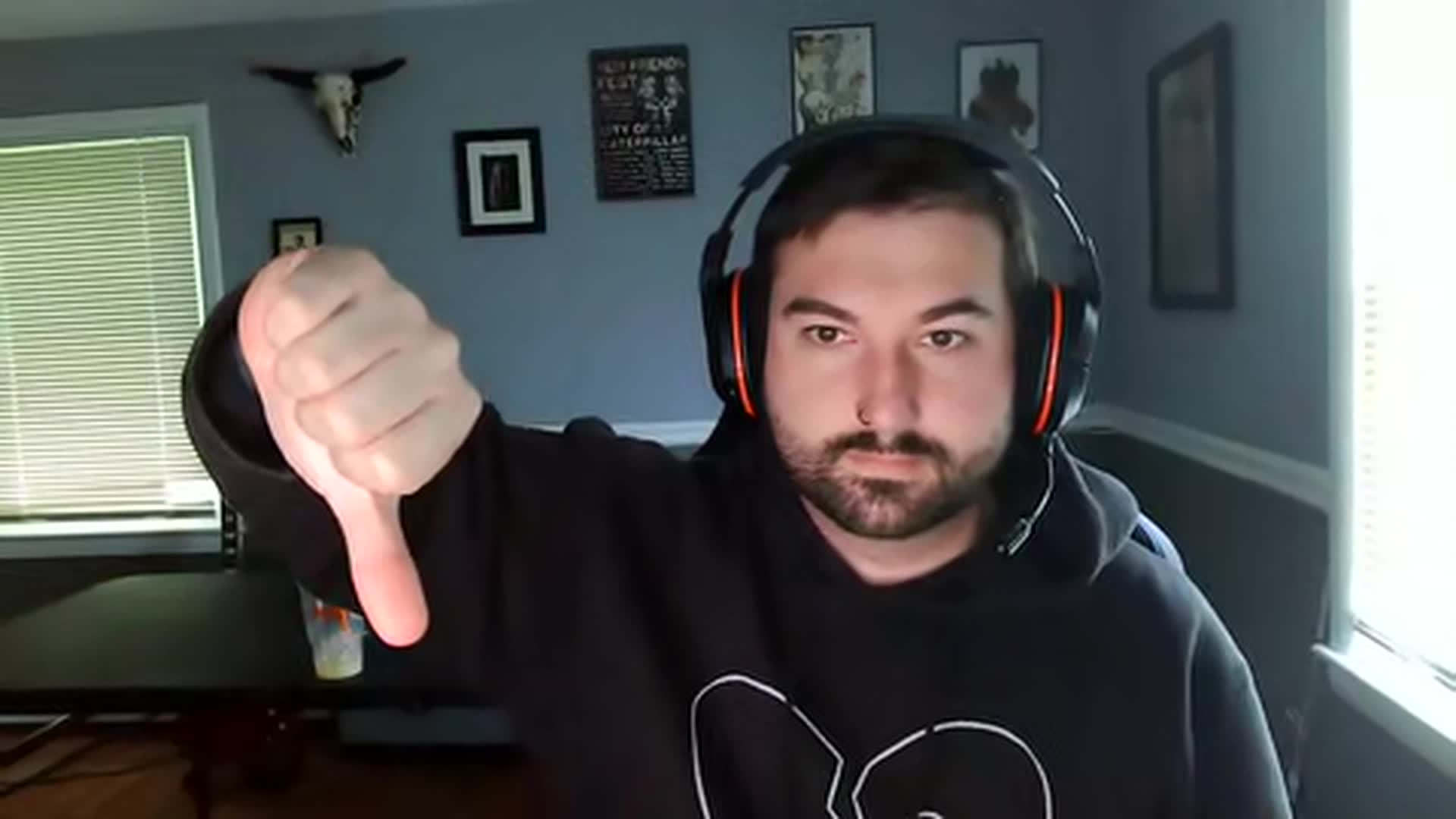}}
        \text{Sequence2 Bicubic}\vspace{.1cm}
            \end{minipage}
    \begin{minipage}[b]{0.325\linewidth}
        \centering
        \centerline{\includegraphics[width=.98\linewidth]{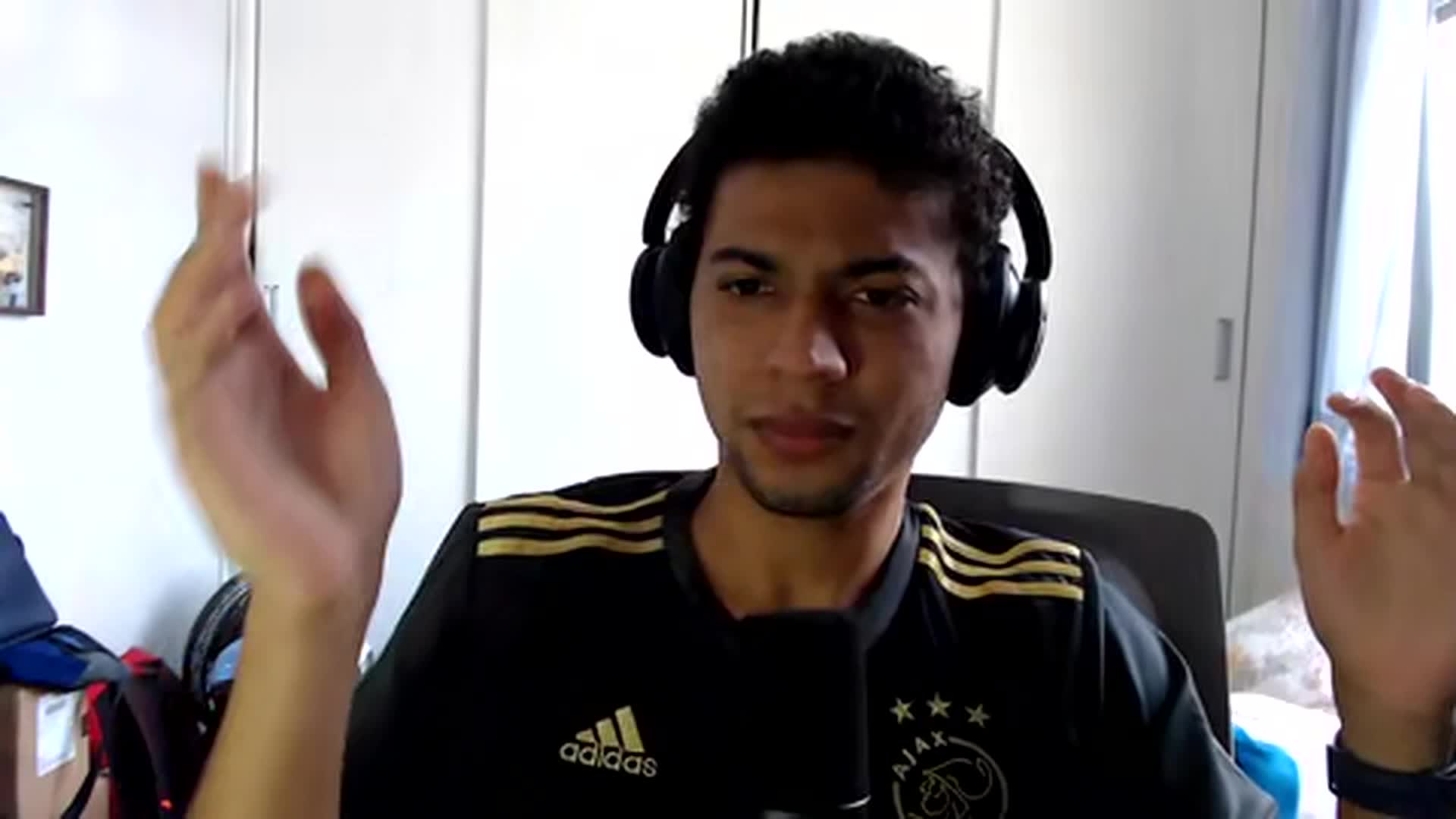}}
        \text{Sequence3 Bicubic}\vspace{.1cm}
            \end{minipage}

    \begin{minipage}[b]{0.325\linewidth}
        \centering
        \centerline{\includegraphics[width=.98\linewidth]{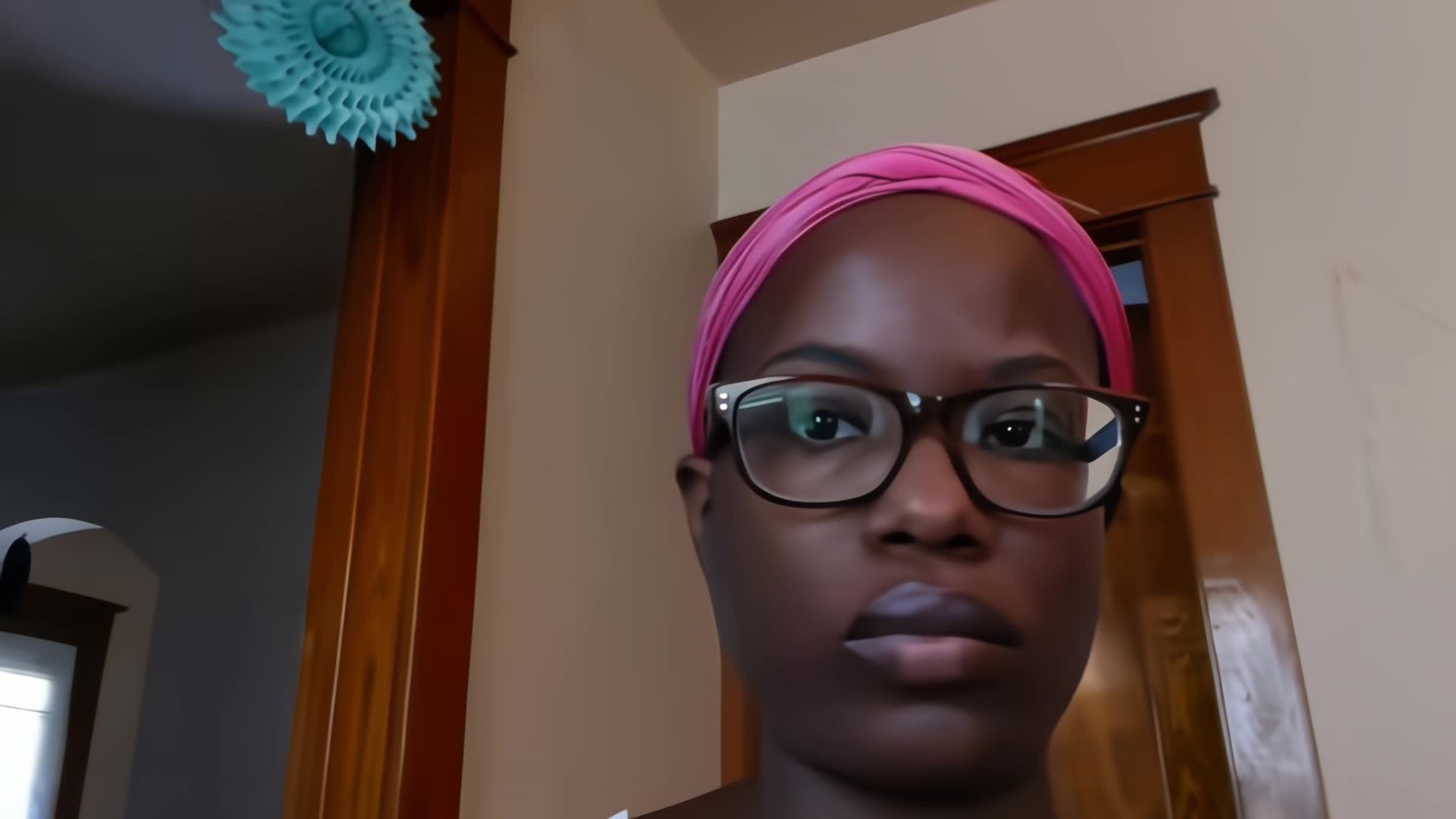}}
        \text{Sequence1 Ours}\vspace{.1cm}
        \end{minipage}
    \begin{minipage}[b]{0.325\linewidth}
        \centering
        \centerline{\includegraphics[width=.98\linewidth]{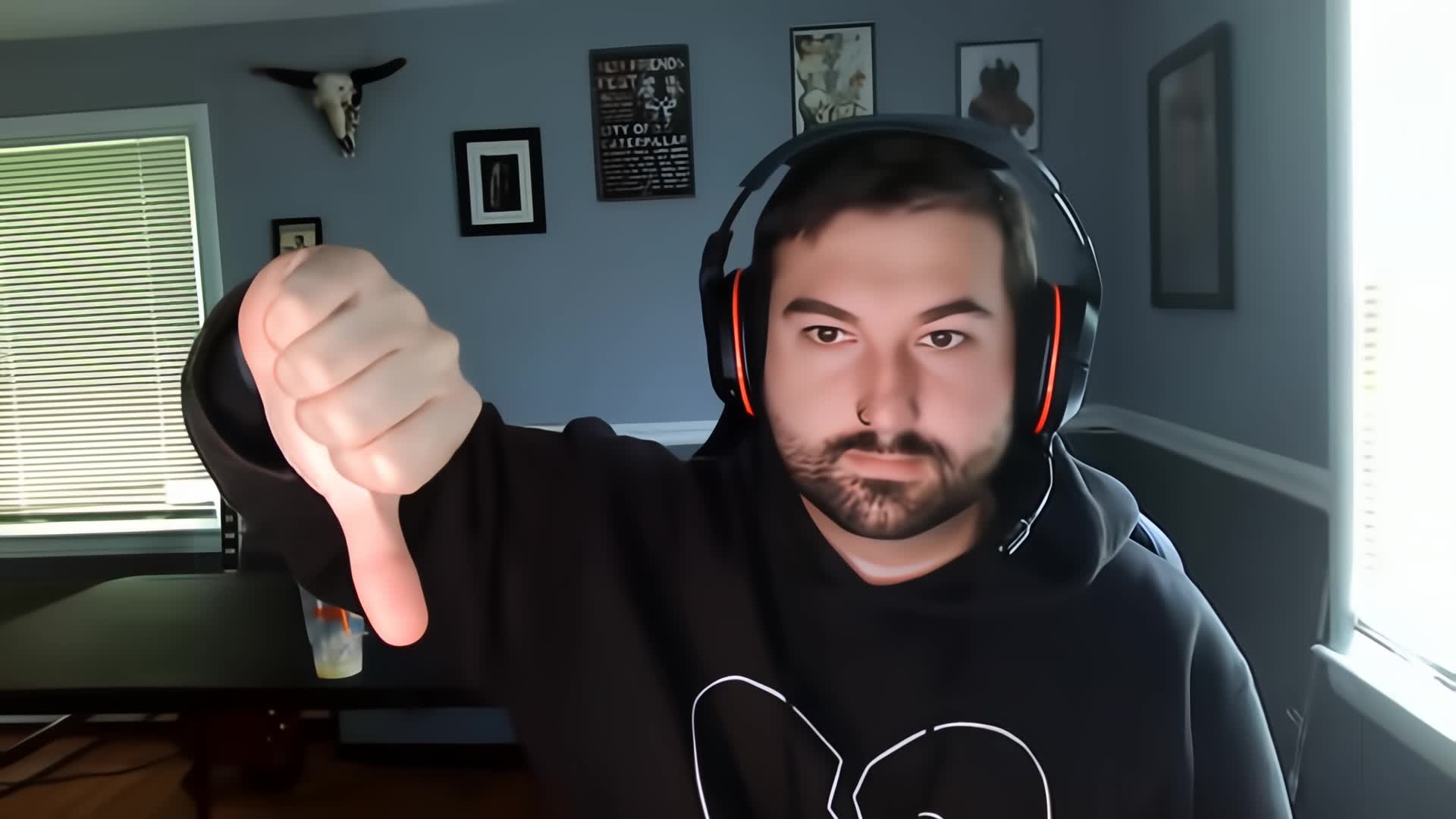}}
        \text{Sequence2 Ours}\vspace{.1cm}
            \end{minipage}
    \begin{minipage}[b]{0.325\linewidth}
        \centering
        \centerline{\includegraphics[width=.98\linewidth]{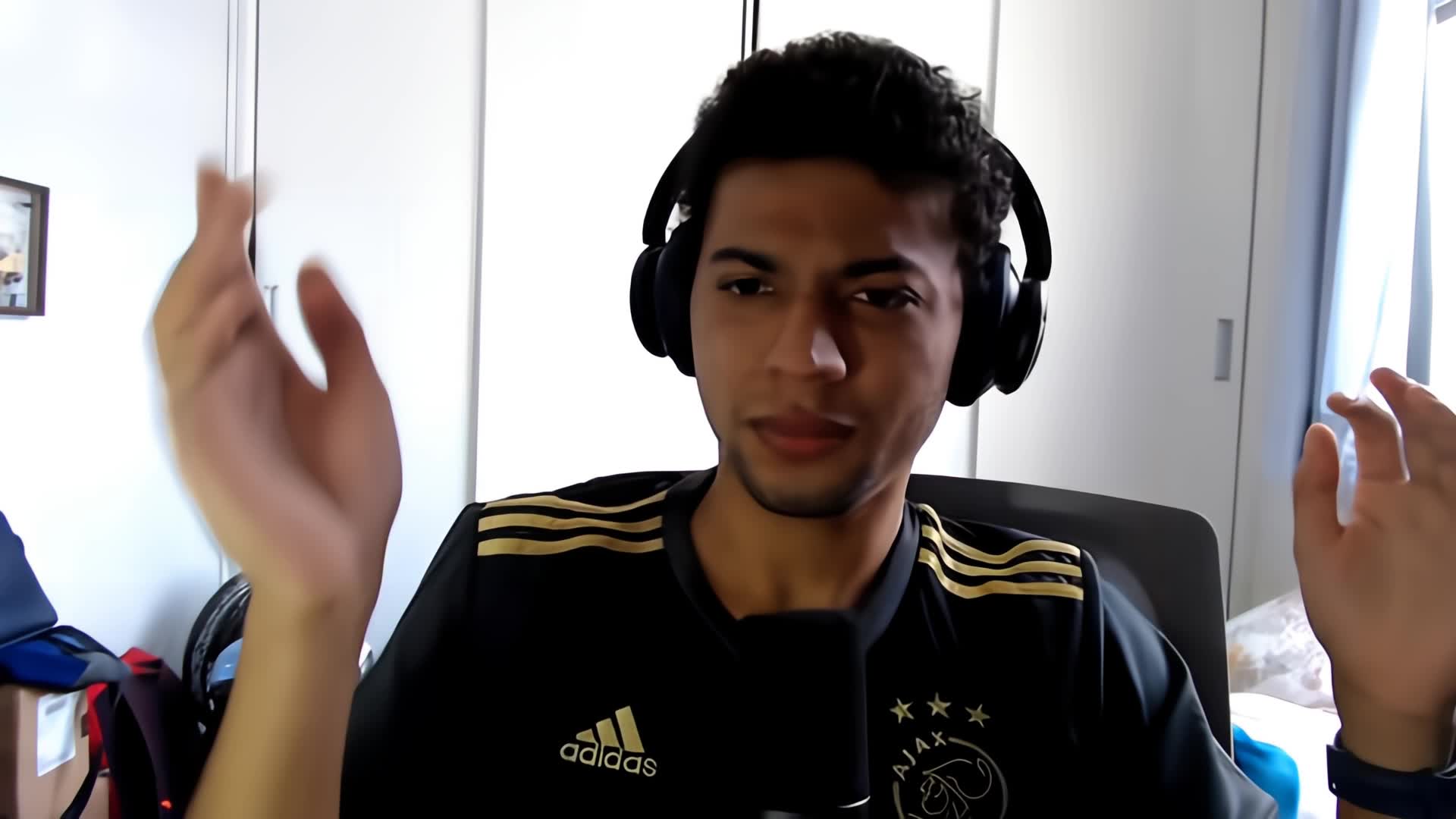}}
        \text{Sequence3 Ours}\vspace{.1cm}
            \end{minipage}
    
    
    \caption{Visual comparison of track2 SR reconstruction results.}
    \label{fig:example2}
\end{figure*}

\subsection{Employed Network Architecture}

The overall architecture of the proposed model is depicted in \autoref{fig:netwoork}. Specifically, a compressed 64$\times$64 YCbCr 4:2:0 image block is first processed by a nearest-neighbor (NN) upsampling operation to restore its chroma channels, resulting in a 64$\times$64 YCbCr 4:4:4 input. This preprocessed block is then fed into the super-resolution network, which is designed to predict a high-resolution 256$\times$256 YCbCr 4:4:4 image block, achieving a spatial upscaling factor of $4\times$. To ensure compatibility with standard coding pipelines, the network output is subsequently converted back to the YCbCr 4:2:0 format.

The network backbone is constructed based on the recently proposed HiET (Hierarchical Encoding Transformer) layer \cite{jiang2025c2d}, which efficiently captures both local spatial structures and long-range contextual dependencies. Building upon this design, we propose a refined network architecture specifically optimized for compressed-domain restoration and super-resolution tasks. As shown in \autoref{fig:netwoork}, the HiET layers are configured with window sizes of [64, 32, 8, 32, 64], where the number of stacked blocks is set to $B=6$ and the hidden channel dimension is fixed at 126. This configuration is carefully selected to balance model capacity and computational efficiency, making it particularly suitable for practical deployment in video conferencing scenarios.

\subsection{Training Configuration}

The training process of the proposed VSR-HE model is divided into two stages.

In the first stage, the network is optimized using a combined perceptual loss function based on \cite{ma2020cvegan}, which balances pixel-wise accuracy and perceptual fidelity:
\begin{equation}
    \mathcal{L}_{p} = 0.3 \mathcal{L}_\mathit{L1} + 0.2 \mathcal{L}_\mathit{SSIM} + 0.1 \mathcal{L}_\mathit{L2} + 0.4 \mathcal{L}_\mathit{MS-SSIM}
\end{equation}
where $\mathcal{L}_\mathit{L1}$ and $\mathcal{L}_\mathit{L2}$ denote the pixel-wise L1 and L2 losses, respectively, while $\mathcal{L}_\mathit{SSIM}$ and $\mathcal{L}_\mathit{MS-SSIM}$ represent the Structural Similarity Index and its multi-scale variant. This combined objective ensures both structural preservation and enhanced perceptual quality during the early training phase.

In the second stage, following the strategy proposed in \cite{wang2018esrgan}, we further introduce an adversarial loss component based on the GAN framework to refine the perceptual realism of the super-resolved outputs. The total loss in this stage is formulated as the weighted sum of the perceptual loss $\mathcal{L}_{p}$ and the GAN loss $\mathcal{L}_{GAN}$.
\begin{equation}
    \mathcal{L}_{total} = \mathcal{L}_{p} + 0.05\mathcal{L}_{GAN}.
\end{equation}

The employed model is implemented using PyTorch 1.10 \cite{paszke2019pytorch}. Training is performed with the Adam optimizer \cite{kingma2014adam}, with default hyperparameters $\beta_{1} = 0.9$ and $\beta_{2} = 0.999$. A batch size of 16 is used. The learning rate is initially set to $1\times10^{-4}$ and is progressively reduced by a factor of 2 at 50k, 100k, 200k, and 300k iterations, consistently across both training stages to facilitate stable convergence. Both training and evaluation were executed on an NVIDIA RTX4090 GPU.

\subsection{Training Content}

The proposed VSR-HE model is optimized through supervised training on a curated dataset, as detailed below. In addition to the provided REDS dataset \cite{nah2019ntire} for Track 1 and the VCD dataset \cite{naderi2024vcd} for Track 2, we further incorporate original video sequences from the BVI-AOM database \cite{nawala2024bvi} to diversify and enhance the training corpus. Representative thumbnails from the dataset are illustrated in \autoref{fig:bviexample}.

\begin{table*}[t]
\caption{PSNR-Y, SSIM, MS-SSIM, and VMAF results for the proposed methods and all benchmarks for both Track 1 and 2.}
\centering
\resizebox{1\linewidth}{!}{ \begin{tabular}{r c  c c  c c c c c  c}
\toprule
 & \multicolumn{4}{c}{Track1}  & \multicolumn{4}{c}{Track2}  \\
 \cmidrule{2-5} \cmidrule{7-10}
 Method & PSNR-Y (dB)$\uparrow$ & SSIM$\uparrow$ & MS-SSIM$\uparrow$ & VMAF$\uparrow$   && PSNR-Y (dB)$\uparrow$ & SSIM$\uparrow$ & MS-SSIM$\uparrow$ & VMAF$\uparrow$  \\ \midrule
   Bicubic & 25.74  & 0.8425 &  0.8538 & 37.87 &&  - & - & -  & - \\
   Lanczos & -  & - &  - & - &&  32.64 & 0.9612 & 0.9611  & 40.97\\
 EDSR \cite{lim2017enhanced} & 26.11  & 0.8710 & 0.8745  & 50.34 && 32.96  & 0.9611 & 0.9613  & 53.64\\
 CVEGAN \cite{ma2020cvegan} & 26.15 & 0.8751 &  0.8801 & 58.31 && 32.92  & 0.9621 &  0.9619 & 59.49\\
 SwinIR \cite{liang2021swinir} &  26.40 & 0.8711 & 0.8788  & 56.87 &&  33.12 & 0.9612 & 0.9620  & 57.56\\
  \textbf{Ours}   & \textbf{26.47} & \textbf{0.8759} &  \textbf{0.8829} & \textbf{59.17} && \textbf{33.17}  & \textbf{0.9635} &  \textbf{0.9650} & \textbf{60.12} \\ 
  \bottomrule
\end{tabular}}

    \label{tab:performance}
\end{table*}

\begin{table*}[t]
\caption{Model complexity results.}
    \centering
    \resizebox{0.85\linewidth}{!}{\begin{tabular}{cc c c c c c c}
    \toprule
         Track & Input & Output& Train Time (hrs) & \# Params. (M) & FLOPs (G) & GPU & Runtime (ms/frame)  \\
        \midrule
        1& 180p & 720p & 240 & 5.43 & 455.16 & RTX4090 & 140.61 \\
         2& 270p & 1080p & 240 & 5.43 & 455.16 & RTX4090 & 375.11 \\
         \bottomrule
    \end{tabular}}

    \label{tab:my_label}
\end{table*}

All supplemental sequences were encoded by HEVC HM 18.0 with five different quantization parameter (QP) values: 17, 22, 27, 32, 34, and 37, after downsampling. Following \cite{ma2020cvegan}, both the degraded sequences and their corresponding high-quality originals were uniformly cropped into 64$\times$64  (compressed input) and 256$\times$256 patches (high-resolution ground truth), respectively. These patches were randomly sampled to construct the training pairs. To further enhance data diversity and improve generalization, common augmentation techniques such as random rotations and horizontal/vertical flipping were applied. This comprehensive data preparation process resulted in approximately 100,000 patch pairs for each track. The model was trained independently on the respective datasets for Track 1 and Track 2, enabling it to effectively handle compressed video content across a broad range of QP values while maintaining robustness to various compression artifacts.

\section{Results and Discussion}
\label{sec:RD}
Five sequences, provided by the ICME 2025 grand challenge organizer, are used to evaluate the effectiveness of the proposed coding framework. Each sequence contains up to 300 frames and is compressed with six different QPs, ranging from 17 to 37, after down-sampling. The decoded sequences were also provided by the organizer and stored in mp4 format. These sequences were first converted into YCbCr 4:4:4 format and then input into the VSR-HE model to recover to their original resolution.

\autoref{tab:performance} summarizes the average performance of the proposed VSR-HE method for the test sequences in terms of VMAF, SSIM, MS-SSIM and PSNR-Y for both tracks. To benchmark the performance of our proposed VSR-HE, we also test several other methods, including bicubic filter, EDSR \cite{lim2017enhanced}, CVEGAN \cite{ma2020cvegan} and SwinIR \cite{liang2021swinir}. According to the evaluation results, the proposed model demonstrates strong performance advantages across multiple aspects, including perceptual quality and fidelity to the original content. Visual comparisons with Bicubic/Lanczos filters are presented in \autoref{fig:example1} for Track 1 and \autoref{fig:example2} for Track 2. As shown, the proposed VSR-HE model effectively mitigates compression artifacts and reconstructs finer image details. These results highlight the model’s effectiveness in enhancing real-world compressed videos and its potential for various video enhancement applications in practical scenarios. 

Moreover, we also report the training time, number of parameters, FLOPs, and runtime in \autoref{tab:my_label}. As shown in the table, the total number of parameters of VSR-HE is 5.43M and the processing speed for each frame is 140 ms. These results offer valuable insights for the organizers to conduct an in-depth analysis and comparison of the strengths and weaknesses of each participating method.

\section{Conclusion}
\label{sec:c}

In this paper, we propose VSR-HE, a super-resolution framework designed to upscale 180p compressed videos to 720p resolution. The proposed method has been tested on H.265/HEVC compressed content and submitted to the ICME 2025 Grand Challenge on Video Super-Resolution for Video Conferencing, Track 1: General-Purpose Real-World Video Content with $4\times$ Upscaling (Team BVI-VSR). The evaluation results demonstrate that VSR-HE can significantly enhance visual quality while maintaining compatibility with existing video coding workflows. Future work will aim to further improve the computational efficiency of the model and extend its deployment to other standard codecs and application scenarios.

\section*{Acknowledgment}
The authors appreciate the funding from the University of Bristol, the UKRI MyWorld Strength in Places Programme (SIPF00006/1), and the China Scholarship Council.

\small

\bibliographystyle{ieeetr}
\bibliography{refs}

\begin{thebibliography}{10}

\bibitem{r:cisco2}
CISCO, ``{CISCO} visual networking index: forecast and methodology, 2017--2022,'' November 2018.

\bibitem{h265HEVC}
G.~J. Sullivan, J.-R. Ohm, W.-J. Han, and T.~Wiegand, ``{Overview of the High Efficiency Video Coding (HEVC) Standard},'' {\em IEEE Transactions on Circuits and Systems for Video Technology}, vol.~22, no.~12, pp.~1649--1668, 2012.

\bibitem{VVCSoftware_VTM}
``{VVCSoftware\_VTM}.'' \url{https://vcgit.hhi.fraunhofer.de/jvet/VVCSoftware\_VTM}.
\newblock Accessed: Enter Date Accessed.

\bibitem{AV1}
``{SVT-AV1}.'' \url{https://gitlab.com/AOMediaCodec/SVT-AV1}.
\newblock Accessed: Enter Date Accessed.

\bibitem{VP9}
``{VP9}.'' \url{https://www.webmproject.org/vp9/}.
\newblock Accessed: Enter Date Accessed.

\bibitem{ecm_github}
{Joint Video Experts Team (JVET)}, ``{Enhanced Compression Model (ECM) 12.0 Library }.'' \url{https://vcgit.hhi.fraunhofer.de/ecm/ECM}, 2024.
\newblock Accessed: 2024-04-10.

\bibitem{avm_github}
{Alliance for Open Media}, ``{AOM Video Model (AVM) Codec 2.0.0 Library}.'' \url{https://gitlab.com/AOMediaCodec/avm}, 2024.
\newblock Accessed: 2024-04-10.

\bibitem{teng2024benchmarking}
S.~Teng, Y.~Jiang, G.~Gao, F.~Zhang, T.~Davis, Z.~Liu, and D.~Bull, ``Benchmarking conventional and learned video codecs with a low-delay configuration,'' in {\em 2024 IEEE International Conference on Visual Communications and Image Processing (VCIP)}, pp.~1--5, IEEE, 2024.

\bibitem{jiang2024mtkd}
Y.~Jiang, C.~Feng, F.~Zhang, and D.~Bull, ``{MTKD}: Multi-teacher knowledge distillation for image super-resolution,'' {\em arXiv preprint arXiv:2404.09571}, 2024.

\bibitem{liang2021swinir}
J.~Liang, J.~Cao, G.~Sun, K.~Zhang, L.~Van~Gool, and R.~Timofte, ``{SwinIR}: Image restoration using swin transformer,'' in {\em Proceedings of the IEEE/CVF international conference on computer vision}, pp.~1833--1844, 2021.

\bibitem{jiang2024hiif}
Y.~Jiang, H.~M. Kwan, T.~Peng, G.~Gao, F.~Zhang, X.~Zhu, J.~Sole, and D.~Bull, ``Hiif: Hierarchical encoding based implicit image function for continuous super-resolution,'' {\em arXiv preprint arXiv:2412.03748}, 2024.

\bibitem{jiang2025c2d}
Y.~Jiang, C.~Zeng, S.~Teng, F.~Zhang, X.~Zhu, J.~Sole, and D.~Bull, ``C2d-isr: Optimizing attention-based image super-resolution from continuous to discrete scales,'' {\em arXiv preprint arXiv:2503.13740}, 2025.

\bibitem{zhu2025blind}
Q.~Zhu, Y.~Jiang, S.~Zhu, F.~Zhang, D.~Bull, and B.~Zeng, ``Blind video super-resolution based on implicit kernels,'' {\em arXiv preprint arXiv:2503.07856}, 2025.

\bibitem{lim2017enhanced}
B.~Lim, S.~Son, H.~Kim, S.~Nah, and K.~Mu~Lee, ``Enhanced deep residual networks for single image super-resolution,'' in {\em Proceedings of the IEEE conference on computer vision and pattern recognition workshops}, pp.~136--144, 2017.

\bibitem{peng2025instance}
T.~Peng, H.~M. Kwan, Y.~Jiang, G.~Gao, F.~Zhang, X.~Xu, S.~Liu, and D.~Bull, ``Instance data condensation for image super-resolution,'' {\em arXiv preprint arXiv:2505.21099}, 2025.

\bibitem{yan2018convolutional}
N.~Yan, D.~Liu, H.~Li, B.~Li, L.~Li, and F.~Wu, ``Convolutional neural network-based fractional-pixel motion compensation,'' {\em IEEE Transactions on Circuits and Systems for Video Technology}, vol.~29, no.~3, pp.~840--853, 2018.

\bibitem{zhang2020enhancing}
F.~Zhang, C.~Feng, and D.~R. Bull, ``Enhancing vvc through cnn-based post-processing,'' in {\em 2020 IEEE International Conference on Multimedia and Expo (ICME)}, pp.~1--6, IEEE, 2020.

\bibitem{ma2020cvegan}
D.~Ma, F.~Zhang, and D.~R. Bull, ``{CVEGAN}: a perceptually-inspired gan for compressed video enhancement,'' {\em arXiv preprint arXiv:2011.09190}, 2020.

\bibitem{ma2020mfrnet}
D.~Ma, F.~Zhang, and D.~R. Bull, ``{MFRNet}: a new {CNN} architecture for post-processing and in-loop filtering,'' {\em IEEE Journal of Selected Topics in Signal Processing}, vol.~15, no.~2, pp.~378--387, 2020.

\bibitem{jiang2024rtsr}
Y.~Jiang, J.~Nawa{\l}a, C.~Feng, F.~Zhang, X.~Zhu, J.~Sole, and D.~Bull, ``Rtsr: A real-time super-resolution model for av1 compressed content,'' {\em arXiv preprint arXiv:2411.13362}, 2024.

\bibitem{jiang2024compressing}
Y.~Jiang, J.~Nawa{\l}a, F.~Zhang, and D.~Bull, ``Compressing deep image super-resolution models,'' in {\em 2024 Picture Coding Symposium (PCS)}, pp.~1--5, IEEE, 2024.

\bibitem{feng2022vistra3}
C.~Feng, D.~Danier, C.~Tan, F.~Zhang, and D.~Bull, ``Vistra3: Video coding with deep parameter adaptation and post processing,'' in {\em 2022 IEEE International Symposium on Circuits and Systems (ISCAS)}, pp.~824--828, IEEE, 2022.

\bibitem{conde2024aim}
M.~V. Conde, Z.~Lei, W.~Li, C.~Bampis, I.~Katsavounidis, and R.~Timofte, ``Aim 2024 challenge on efficient video super-resolution for av1 compressed content,'' {\em arXiv preprint arXiv:2409.17256}, 2024.

\bibitem{zhu2024cpga}
Q.~Zhu, J.~Hao, Y.~Ding, Y.~Liu, Q.~Mo, M.~Sun, C.~Zhou, and S.~Zhu, ``Cpga: Coding priors-guided aggregation network for compressed video quality enhancement,'' in {\em Proceedings of the IEEE/CVF Conference on Computer Vision and Pattern Recognition}, pp.~2964--2974, 2024.

\bibitem{zhu2025fcvsr}
Q.~Zhu, F.~Zhang, F.~Chen, S.~Zhu, D.~Bull, and B.~Zeng, ``Fcvsr: A frequency-aware method for compressed video super-resolution,'' {\em arXiv preprint arXiv:2502.06431}, 2025.

\bibitem{nawala2024bvi}
J.~Nawa{\l}a, Y.~Jiang, F.~Zhang, X.~Zhu, J.~Sole, and D.~Bull, ``{BVI-AOM}: A new training dataset for deep video compression optimization,'' {\em arXiv preprint arXiv:2408.03265}, 2024.

\bibitem{wang2018esrgan}
X.~Wang, K.~Yu, S.~Wu, J.~Gu, Y.~Liu, C.~Dong, Y.~Qiao, and C.~Change~Loy, ``Esrgan: Enhanced super-resolution generative adversarial networks,'' in {\em Proceedings of the European conference on computer vision (ECCV) workshops}, pp.~0--0, 2018.

\bibitem{paszke2019pytorch}
A.~Paszke, S.~Gross, F.~Massa, A.~Lerer, J.~Bradbury, G.~Chanan, T.~Killeen, Z.~Lin, N.~Gimelshein, L.~Antiga, {\em et~al.}, ``Pytorch: An imperative style, high-performance deep learning library,'' {\em Advances in neural information processing systems}, vol.~32, 2019.

\bibitem{kingma2014adam}
D.~P. Kingma and J.~Ba, ``Adam: A method for stochastic optimization,'' {\em arXiv preprint arXiv:1412.6980}, 2014.

\bibitem{nah2019ntire}
S.~Nah, S.~Baik, S.~Hong, G.~Moon, S.~Son, R.~Timofte, and K.~Mu~Lee, ``Ntire 2019 challenge on video deblurring and super-resolution: Dataset and study,'' in {\em Proceedings of the IEEE/CVF conference on computer vision and pattern recognition workshops}, pp.~0--0, 2019.

\bibitem{naderi2024vcd}
B.~Naderi, R.~Cutler, N.~S. Khongbantabam, Y.~Hosseinkashi, H.~Turbell, A.~Sadovnikov, and Q.~Zou, ``Vcd: A video conferencing dataset for video compression,'' in {\em ICASSP 2024-2024 IEEE International Conference on Acoustics, Speech and Signal Processing (ICASSP)}, pp.~3970--3974, IEEE, 2024.

\end{thebibliography}

\end{document}